\documentclass[final,3p,times,10.5pt]{elsarticle}
\usepackage[latin1,utf8]{inputenc}
\usepackage{amssymb}
\usepackage{amsmath}
\usepackage{dsfont}
\usepackage{float}
\usepackage{graphicx}
\usepackage[labelformat=parens,font=scriptsize,position=t,singlelinecheck=off,topadjust,margin=18pt,captionskip=-5pt]{subfig}
\usepackage{caption}
\usepackage[usenames,dvipsnames]{color}
\usepackage{hyperref}
\hypersetup{colorlinks=true}
\graphicspath{{./figs/pdf/}}
\DeclareGraphicsExtensions{.eps,.pdf}

\newcommand{\vect}[1]{\boldsymbol{#1}}
\newcommand{\mat}[1]{\boldsymbol{#1}}
\biboptions{numbers,square}
\journal{Physica A: Statistical Mechanics and its Applications.}
\begin{document}
\begin{frontmatter}
\title{Maximum approximate entropy and r threshold: a new approach for regularity changes detection}
\author[l1]{Juan F. Restrepo}
\author[l1,l2]{Gastón Schlotthauer}
\author[l1,l2]{María E. Torres\corref{c2}}
\ead{metorres@santafe-conicet.gov.ar}
\address[l1]{Laboratorio de Señales y Dinámicas no Lineales,  Facultad de Ingeniería, Universidad Nacional de Entre Ríos,\\
Ruta Prov. 11 Km. 10. Oro Verde - Entre Ríos, Argentina}
\address[l2]{National Scientific and Technical Research Council (CONICET), Argentina}
\cortext[c1]{Published version: \href{http://dx.doi.org/10.1016/j.physa.2014.04.041}{http://dx.doi.org/10.1016/j.physa.2014.04.041}}
\cortext[c2]{Corresponding author. Tel: +54-(0)343-4975100 (122)}
\begin{abstract}
Approximate entropy (\textit{ApEn})  has been widely used as  an estimator of regularity in  many scientific fields.  It
has proved  to be a useful  tool because of its ability  to distinguish different  system's dynamics when  there is only
available short-length  noisy data.  Incorrect  parameter selection  (embedding dimension  $m$,  threshold $r$  and data
length $N$) and  the presence of noise in  the signal can undermine the  \textit{ApEn} discrimination capacity.  In this
work we show that $r_{max}$ ($ApEn(m,r_{max},N)=ApEn_{max}$) can also  be used as a feature to discern between dynamics.
Moreover,  the combined use  of $ApEn_{max}$ and $r_{max}$  allows a better discrimination capacity  to be accomplished,
even in  the presence of  noise.  We conducted our  studies using real  physiological time series  and simulated signals
corresponding to both low- and high-dimensional systems.  When $ApEn_{max}$ is incapable of discerning between different
dynamics because of the noise presence,  our results suggest  that $r_{max}$ provides additional information that can be
useful for classification purposes.  Based on cross-validation tests,  we conclude that, for short length noisy signals,
the  joint use  of $ApEn_{max}$  and $r_{max}$  can  significantly  decrease  the  misclassification  rate  of  a linear
classifier in comparison with their isolated use.
\end{abstract}
\end{frontmatter}
\section{Introduction}
\label{int}
The concept of changing complexity has proved to be helpful to characterize and assess different phenomena in areas such
as seismology, economy, mechanics, physiology, etc.  \cite{potirakis2012,isik2010,He2012,Vaillancourt2002}.  In the last
30 years  this challenging  endeavor has led  researchers and practitioners  to develop  different methods  conceived to
estimate and understand such complexity changes and their relationship with physical and biological system dynamics.  In
the early nineties,  Lipsitz et al.  reported that the process of  natural aging is attached to a decrease of complexity
in the dynamics of physiological functions \cite{Lipsitz1992a}.  This results  in a loss in the capacity of the organism
to adapt to stress, making it more vulnerable to diseases.

Approximate entropy finds its  origins in Kolmogorov-Sinai Entropy (\textit{K-S Entropy}),  defined as  the mean rate of
information generated by  a  process  \cite{Kol1958,Eckmann1985}.  This  measure  is  recognized  for being a meaningful
parameter to describe  the behavior of dynamical systems.  In \cite{Grassberger1983}  Grassberger and Procaccia provided
an  algorithm  to  calculate  a  lower  bound  for   the  \textit{K-S  Entropy}  from  a  finite  time  series.   Takens
\cite{Takens1983} and Eckmann and Ruelle \cite{Eckmann1985} modified  this approach to directly evaluate the \textit{K-S
Entropy}.   Motivated  by  \textit{Eckmann-Ruelle  Entropy},  Pincus  introduced  the  \textit{ApEn}  \cite{Pincus1991},
providing  a  statistic  to  assess complexity  from  noisy  short-length  data.  For  an  $N$-dimensional  time series,
\textit{ApEn} depends on  two  parameters:  the  \textit{Embedding  Dimension}  ($m$)  and the \textit{Threshold} ($r$).
$ApEn(m,r)$ and $ApEn(m,r,N)$ can be seen respectively as a family of parametric statistics and estimators,  designed to
measure the regularity  of a system.  \textit{ApEn} has been  widely used as a non-linear  feature to classify different
dynamics, for example epileptic seizures \cite{Acharya2012a,Srinivasan2007,Shen2011} and sleep apnea \cite{Acharya2011}.

Because   of   the   bias   introduced   by   counting   self-matches   and   the   finite   data   length   ($N$),   in
\cite{Pincus1994,Richman2000a} the  authors assert that the  estimator $ApEn(m,r,N)$ lacks  of consistency.  To overcome
this limitation,  Richman et al.  proposed the Sample Entropy  (\textit{SampEn}) as a more consistent regularity measure
\cite{Richman2000a}.  However,  both  \textit{ApEn} and  \textit{SampEn} are  highly  dependent  on  the  set  of chosen
parameters ($m$,  $r$).  Chon et al.  \cite{Chon2009} assert that  neither \textit{ApEn} nor \textit{SampEn} is accurate
in measuring the signal's  complexity when the calculations are made  with the values of $m$ and  $r$ recommended in the
literature \cite{Pincus1991e}.  Instead,  the use of $ApEn_{max}$,  i.e.  the maximum value of $ApEn(m,r,N)$, with fixed
$m$ and $N$ was proposed as a more consistent estimator of system's complexity \cite{Chon2009,Chen2005,Lu2008}.

The signal's  noise level  has an important  influence on $ApEn(m,r,N)$  estimation and  therefore $ApEn_{max}$  is also
affected.  Pincus asserts  that the reliability in  the calculations could be  seriously undermined when  the {Signal to
Noise  Ratio}  (SNR)  is  below  3~dB   \cite{Pincus1991e}.   To  overcome  this  issue,   some  authors  have  proposed
a pre-processing step,  in which,  techniques such as \textit{Empirical Mode Decomposition} (EMD) \cite{Alam2011,Lv2010}
or \textit{Dyadic Wavelet Transform} (DWT) \cite{Ocak2009} have been used.

In this paper,  we will show that $r_{max}$ ($r$  value at which $ApEn(m,r,N)=ApEn_{max}$) brings useful information and
can be also used as a feature for classification purposes.  Furthermore, the use of $ApEn_{max}$ combined with $r_{max}$
can provide a more consistent method to discern between different dynamics even in presence of noise.

The remainder  of this paper is  organized as follows.  In Section~\ref{sec:met}  we briefly recall  the main approaches
used  for   \textit{ApEn}  parameter  selection   and  we  present   the  methodology  used   in  our  simulations.   In
Section~\ref{sec:res}  the  obtained  results  are  summarized  and  discussed.  Finally,  in  Section~\ref{sec:con} the
conclusions are presented.
\section{Methods}
\label{sec:met}
In order to estimate $ApEn(r,m,N)$ for an $N$-dimensional time series $\left\{u_{1},u_{2},\dots,  u_{N}\right\}$,  given
the   parameters   $m,~\tau~\in~\mathds{N}$,    and   $r\in\mathds{R^{+}}$,   the   $m$-dimensional   embedded   vectors
$\vect{x}_{i}^{m}=\left[u_{i},u_{i+ \tau},u_{i+2\tau},\dots,  u_{i+(m-1)\tau}\right],$  with $1\leq  i\leq N-(m-1)\tau,$
have to be considered.  Then, the \textit{ApEn} is defined as \cite{Pincus1991}:
\[
\begin{array}{@{\hspace{-0.7cm}}l@{\hspace{2cm}}rcl}
      &ApEn\left(m,r,N\right)&=&
      \phi^{m}\left(r\right)-\phi^{m+1}\left(r\right),\\%
     \text{where:}&&&\\%
     &\phi^{m}\left(r\right)&=&
     \frac{1}{N-(m-1)\tau}\sum\limits_{i=1}^{N-(m-1)\tau}{\ln{C_{i}^{m}\left(r\right)}},\\%
     &C_{i}^{m}\left(r\right)&=&
     \frac{1}{N-(m-1)\tau}\sum\limits_{j=1}^{N-(m-1)\tau}{\theta\left(d\left(\vect{x}_{i}^{m},\vect{x}_{j}^{m}\right)
       -r\right)},\\%
     &\theta\left(y\right)&=&
     \left\{\begin{array}{ll} 0 &\quad\text{if}\quad y>0,\\1&\quad\text{otherwise,}\end{array}\right.\\%
     \text{and}&&&\\%
     &d\left(\vect{x}_{i}^{m},\vect{x}_{j}^{m}\right)&=&
     \max\left\{\left| u_{i+k\tau}-u_{j+k\tau} \right|\right\},\qquad 0\leq k \leq m-1.
\end{array}
\]
The \textit{ApEn} measures the logarithmic likelihood that two points ($\vect{x}_{i}^{m}$,  $\vect{x}_{j}^{m}$) that are
close (within  a distance  $r$) in  an $m$-dimensional  space,  remain close  in an  $(m+1)$-dimensional space.  Greater
(lesser)  likelihood of  remaining close  produces  smaller  (larger)  \textit{ApEn}  values  \cite{Pincus1991a}.  It is
important to recall  that $ApEn(m,r)$ was not conceived  as an approximate value of  \textit{E-R Entropy},  therefore it
cannot  certify  chaos.   However,   its  scope   relies  on  its  ability  to  compare   different  types  of  dynamics
\cite{Pincus1991}.  Pincus asserts that,  for a given system,  \textit{ApEn}  values can vary significantly with $m$ and
$r$ \cite{Pincus1991e}.  For this reason, it cannot be seen as an absolute measure.  Moreover, this situation emphasizes
the importance of the  parameters' selection to draw conclusions from \textit{ApEn}  estimations.  In order to make this
paper self contained, we will review some results for parameter selection.

\subsection{Embedding dimension ($m$)} 
\label{ssub:Embedding dimension}
The main purpose of embedding a  time series is to unfold the projection to a state  space that is representative of the
original system's space, i.e.  a reconstructed attractor must preserve the invariant characteristics of the original one
\cite{Faust2012}.  Takens' embedding theorem  gives sufficient conditions to  accomplish this task using  any $m$ bigger
than twice  the \textit{Hausdorff dimension} of  the chaotic attractor.  The idea  is to estimate  the minimum embedding
dimension since a bigger  $m$  will  lead  to  excessive  computational  efforts.  Kennel  et al.  proposed a parametric
algorithm to determine  the minimum embedding dimension,  named \textit{False  Nearest Neighbors} \cite{Kennel992}.  Its
main disadvantage  is that the results  highly depend on the  choice of the algorithm  parameters.  A slightly different
approach was  proposed by  Cao \cite{Cao1997}.  This  method does  not rely  upon subjective  parameters other  than the
embedding lag.

Pincus has suggested to set $m=2$ or  $m=3$ \cite{Pincus1994,Pincus1991e}.  That advice arises from the fact that,  once
$N$  is  set,  high $m$  values  conduct to  poor  $ApEn(m,r)$ estimations.  This  is  due  to  the  bias  introduced by
self-counting and  the decreased number of  vectors $\vect{x}_{i}^{m}$ available  to estimate $C_{i}^{m}\left(r\right)$.
The aforementioned approach may be convenient when low-dimensional systems are studied.  However,  when the dimension is
high,  this criterion  will lead to a  poor reconstruction of the  process' dynamics \cite{Wolf1985,Small2005},  causing
inconsistencies in presence of noise.

It is worthwhile noting that typical applications with  \textit{ApEn} have been conducted using the previously mentioned
values  of $m$.  Aletti  et al.  set  $m=2$ to  assess  congenital  heart  malformation  in  children  using  Heart Rate
Variability  (HRV)  signals  \cite{Aletti2012}.  Zarjam  et  al.  use  $m=2$  and  $3$  to  calculate  $ApEn(m,r)$  from
electroencephalogram (EEG) signals to  investigate changes in working memory load during  the performance of a cognitive
task with varying difficulty levels \cite{Zarjam2012}.

\subsection{Embedding Lag ($\tau$)} 
\label{ssub:Embedding Lag}
The objective of selecting $\tau$ is to maximally spread  the data in the phase space,  removing redundancies and making
fine features more easily  discernible  \cite{Small2005}.  In  most  \textit{ApEn}  applications  $\tau$  is set to one.
Kaffashi et al.  \cite{Kaffashi2008a}  concluded  that,  for  time  series  generated  by  non-linear dynamics and whose
\textit{Autocorrelation  Function} decays  rapidly,  $\tau=1$ is  sufficient to  provide  a  good  estimation  of signal
complexity.  However,  for signals with  long range correlation,  a $\tau$  equal to time occurrence of  the first local
minimum of the \textit{Mutual Information  Function}  or  to  the  time  occurrence  of  the  first zero crossing of the
\textit{Autocorrelation Function} can provide additional information \cite{Small2005,Kaffashi2008a}.

\subsection{Threshold (r)} 
\label{ssub:threshold }
As it was aforementioned,  the statistics $ApEn(m,r)$ can  vary significantly with $r$.  Pincus suggests that $r$ should
lie between  $0.1-0.2$ times the standard  deviation (\textit{SD}) of the  raw signal \cite{Pincus2001,Pincus1992}.  The
$r$ value should be large enough,  not only to avoid significant contribution from noise, but also to admit a reasonable
number of  $\vect{x}_{i}^{m}$ vectors being within  a distance $r$.  This would  ensure an acceptable  estimation of the
$C_{i}^{m}\left(r\right)$ probability \cite{Pincus1991e}.  However,  with too large $r$ values, $ApEn(m,r)$ is unable to
perform fine  process distinctions and consequently,  the  $r$ value selection  will greatly  depend on  the application
\cite{Pincus1994}.

Although the later approach has been broadly applied \cite{Sapoznikov1995a,Pincus1996a,Pincus1996b}, some authors assert
that sometimes  this methodology leads to  an incorrect assessment  of complexity \cite{Chon2009,Chen2005,Lu2008}.  They
proposed the  use of  $ApEn_{max}$ as  a better  complexity estimator.  One  main issue  arises from  the fact  that the
calculation of $ApEn_{max}$ requires  high computational efforts.  To overcome this limitation,  a  set of equations was
proposed to calculate a parameter $\hat{r}_{max}$ as an approximation to $r_{max}$ \cite{Chon2009,Lu2008}.  Supported on
experimental  results with  HRV signals,  Castiglioni  et al.  concluded  that  the  use  of  $ApEn_{max}$  seems  to be
a reasonable  approach,  because this choice would  allow the time series  complexity to be better  quantified than any
other choice of $r$ \cite{Castiglioni2008}.  On the other hand,  Liu et al.  observed that $ApEn_{max}$ was incapable of
distinguishing between groups of healthy and  heart failure subjects,  in experiments with HRV signals.  Further,  since
they found  that $\hat{r}_{max}$ fails in  estimating $r_{max}$ for the  Logistic map,  they asserted that  care must be
taken when using $\hat{r}_{max}$ \cite{Liu2011}.  In a recent study,  Boskovic et al.  \cite{Boskovic2012}, present some
evidence of $ApEn_{max}$ instability.  They observed that for two time series, the estimated $ApEn_{max}$ value suggests
opposite results when data length decreases.  There are other algorithms conceived to reduce the computational effort of
calculating the whole profile of \textit{ApEn} as a function of $m$ and $r$ \cite{Zurek2012,Pan2011}.
\newsavebox\sileq
\begin{lrbox}{\sileq}
  \begin{minipage}{0.15\textwidth}
	\scriptsize
	\begin{align*}
		{dx}/{dt} & = y \\
		{dy}/{dt} & = z \\
		{dz}/{dt} & = \mu x-y-\varepsilon z-ax^{2}-bx^{3}
	\end{align*}
  \end{minipage}
\end{lrbox}

\newsavebox\silp
\begin{lrbox}{\silp}
	\scriptsize
  \begin{minipage}{0.15\textwidth}
	\begin{align*}
		a           & = \left\{0.008,\,\,0.2217\right\}\\
		\varepsilon & = 0.55\\
		\mu         & = 0.65\\
		b           & = 0.65\\
		\Delta t    & = 0.2 \\
	\end{align*}
  \end{minipage}
\end{lrbox}

\newsavebox\silApEnp
\begin{lrbox}{\silApEnp}
	\scriptsize
  \begin{minipage}{0.2\textwidth}
	\begin{align*}
		 m    & = \left\{2,\,\,3,\dots,20\right\}\\
		 r    & = \left\{0,\,\,5\times10^{-5},\dots,0.035\right\}\\
		 \tau & = 10  \\
		 N    & = 5000 \\
			  & \\
	\end{align*}
  \end{minipage}
\end{lrbox}

\newsavebox\maceq
\begin{lrbox}{\maceq}
	\hspace{-3mm}
	\scriptsize
  \begin{minipage}{0.15\textwidth}
	\begin{align*}
		{dx}/{dt}  &= bx + (a\tilde{x})/(1+\tilde{x}^{c}) \\
		\tilde{x}  &= x(t-\Delta)
	\end{align*}
  \end{minipage}
\end{lrbox}

\newsavebox\macp
\begin{lrbox}{\macp}
	\hspace{-6mm}
	\scriptsize
  \begin{minipage}{0.15\textwidth}
	\begin{align*}
		c        & =  \left\{4.5,\,6\right\}\\
		a        & = 0.9\\
		b        & = 0.3\\
		\Delta   & = 80\\
		\Delta t & = 1\\
	\end{align*}
  \end{minipage}
\end{lrbox}

\newsavebox\macApEnp
\begin{lrbox}{\macApEnp}
	\scriptsize
	\renewcommand{\thefootnote}{\alph{footnote}}	
  \begin{minipage}{0.2\textwidth}
	\begin{align*}
		 m    & = \left\{2,\,\,3,\dots,20\right\}\\
		 r    & = \left\{0,\,\,5\times10^{-5},\dots,0.035\right\}\\
		 \tau & = 83  \\
		 N    & = 5000 \\
			  & \\
	\end{align*}
  \end{minipage}
\end{lrbox}

\begin{table}[t!]
  \begin{center}
	\scriptsize
	\begin{tabular}{llll}
                \multicolumn{1}{c}{Model}   &  \multicolumn{1}{c}{Equation}  &   \multicolumn{1}{c}{Model  Parameters}
                & \multicolumn{1}{c}{ApEn Parameters} \\ 
                \hline\hline
		Mackey-Glass     & \usebox{\maceq} & \usebox{\macp} & \usebox{\macApEnp}\\
		\hline
		Shilnikov's type & \usebox{\sileq} & \usebox{\silp} & \usebox{\silApEnp}\\
                \hline\hline
	\end{tabular}
        \caption{Simulation models  and parameters.  $\Delta t$  stands for the time  step used to  obtain the numerical
        solutions.  The upper bound of the $r$ range is equal to the SD of the normalized signal.}
	\label{T1}
  \end{center}
\end{table}
\subsection{Simulations} 
\label{sub:Sim}
In the presence of noise,  the estimator $ApEn_{max}$ could be incapable to discern between different dynamics.  Here we
address the hypothesis that $r_{max}$ provides additional information valuable for the discrimination process.  In other
words,  the  use  of both  $ApEn_{max}$  and $r_{max}$  would  increase  the  ability  of  discerning  between different
complexities in the case of noisy time series.  To assess this hypothesis four simulations were conducted: three of them
with synthetic signals and the last one with an EEG record.

As a  first case  the Mackey-Glass  delay-differential equation  was used  \cite{Cao1997}.  Our aim  is to  assess these
estimators on  time series from a  high-dimensional system \cite{Guevara1983}.  This system  have been used  not only to
study the behavior of complexity estimators on high-dimensions  \cite{Grassberger1983} but also to model the dynamics of
physiological control systems like the neurological system \cite{Guevara1983},  the respiratory system \cite{Mackey1977}
and the hematopoietic system \cite{Mackey1977}.

Two sets of $240$ realizations were produced for each value of the $c$ parameter (see Table~\ref{T1}).  Each realization
with $25000$ points  has a different initial condition,  randomly chosen  from a $\mathcal{U}(0,0.01)$ distribution.  In
order to avoid the influence of transients,  the first $20000$ points of each realization were discarded.  The resulting
signals (with length $N=5000$) were normalized to have unitary energy.  For two randomly selected signals, one from each
set,  its Mutual  Information Function was  calculated.  Then,  the lag corresponding  to each  first local  minimum was
selected, and the $\tau$ parameter was fixed as the largest between these values.  $ApEn(m,r,N)$ was calculated for each
signal,  with $m$ and $r$ taking the values listed in  Table~\ref{T1} and $ApEn_{max}$ and $r_{max}$ were found from the
$ApEn(m,r,N)$ functions.  Additionally an  estimator of the minimum  embedding dimension was calculated  for all signals
using Cao's algorithm \cite{Cao1997}.

With  the  goal  of  analyzing  synthetic data  from  a  system  that  resembles  a  particular  physiological dynamics,
a Shilnikov's type chaos model  was considered as a second case.  The same methodology as  in the first case was adopted
for Shilnikov's  type model  using two values  of the $a$  parameter (see  Table~\ref{T1}),  which allow  simulating EEG
signals  recorded  during a  seizure  of petit  mal  epilepsy  \cite{Friedrich1992}.  The  initial  conditions  for each
realization were selected from a $\mathcal{U}(0,0.01)$ distribution and the $x$ variable was used for the calculations.

In order to evaluate our  method in presence of noise,  white Gaussian noise was added  to each signal (Mackey-Glass and
Shilnikov) with  \textit{SNR}$=5$~dB and \textit{SNR}$=0$~dB.  Then,  all  realizations were normalized  to have unitary
energy  and both  $ApEn_{max}$ and  $r_{max}$ were  calculated as  previously described.  Table~\ref{T1}  summarizes the
models and the  parameter values used to obtain  the time series as well  as the parameter values used  to calculate the
\textit{ApEn}.
\begin{figure}[!t]
	\centering
	\subfloat[][]{\includegraphics[width=0.5\textwidth]{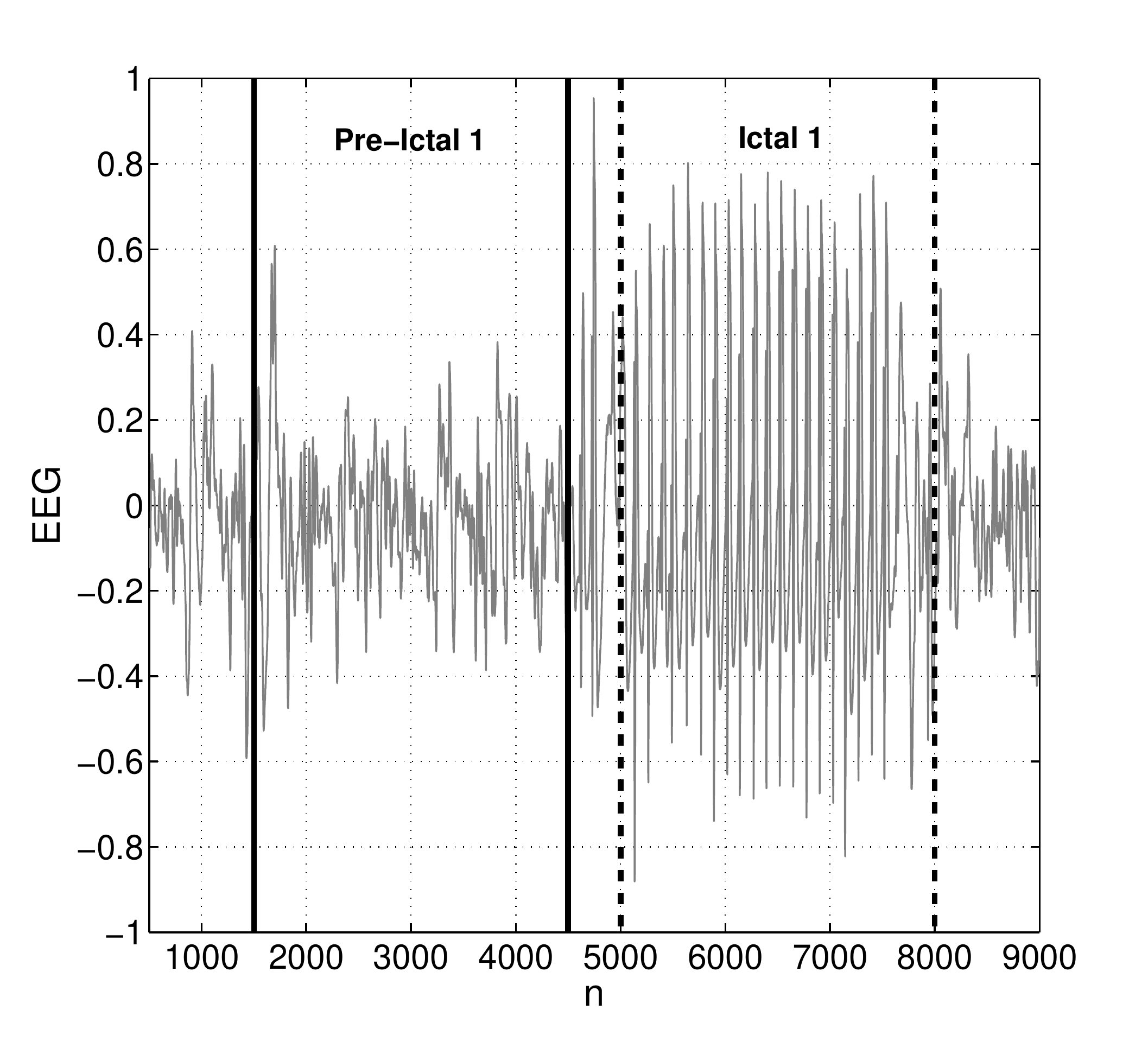}\label{fig:1.1}}
        \subfloat[][]{\includegraphics[width=0.5\textwidth]{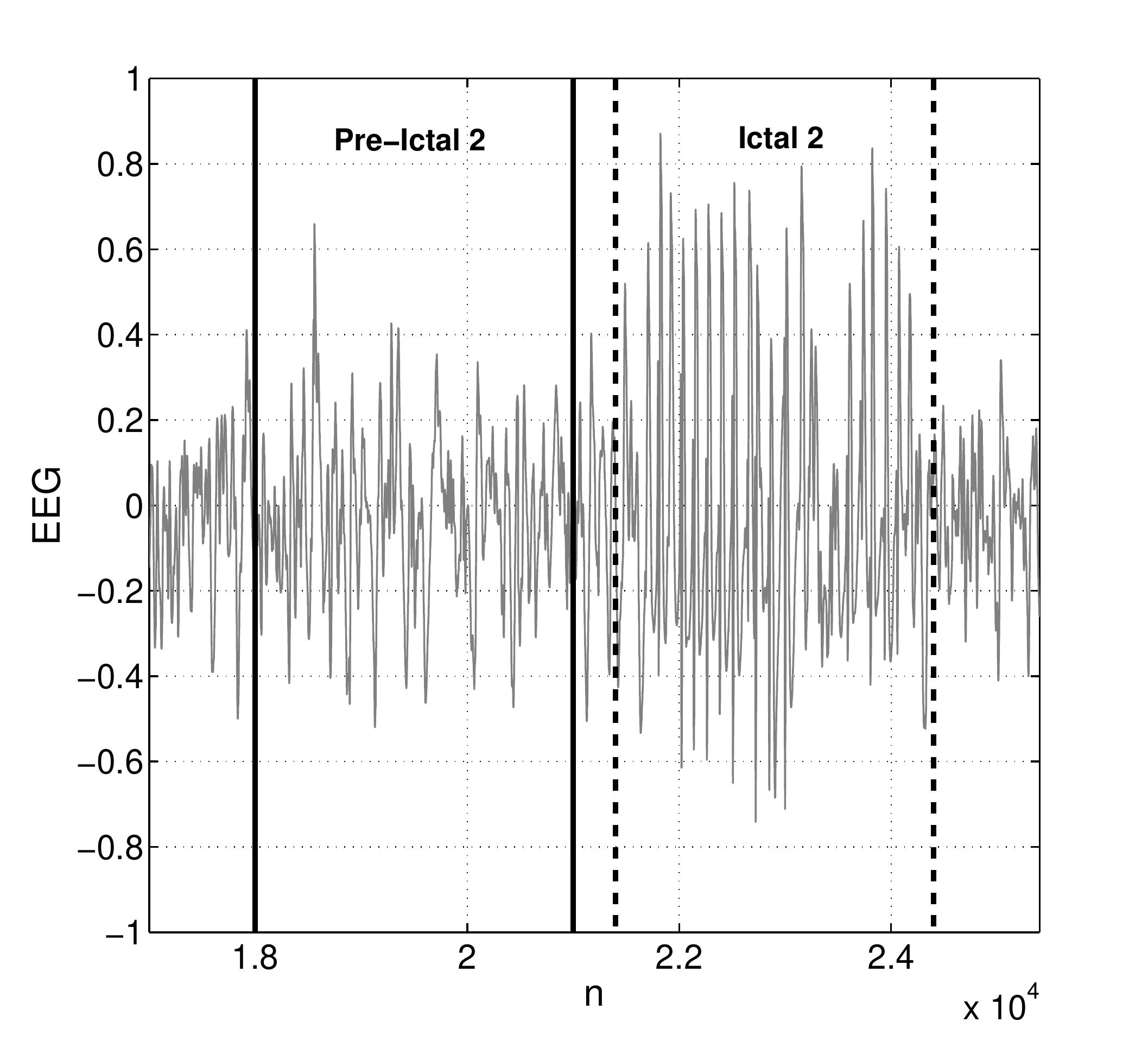}\label{fig:1.2}}
	\caption{EEG signal: (a) First pre-ictal and ictal episodes. (b) Second pre-ictal and ictal episodes.}
	\label{fig:1}
\end{figure}
\begin{figure}[!t]
    \centering
    \subfloat[][]{\includegraphics[width=0.500\linewidth,height=\textheight,keepaspectratio]{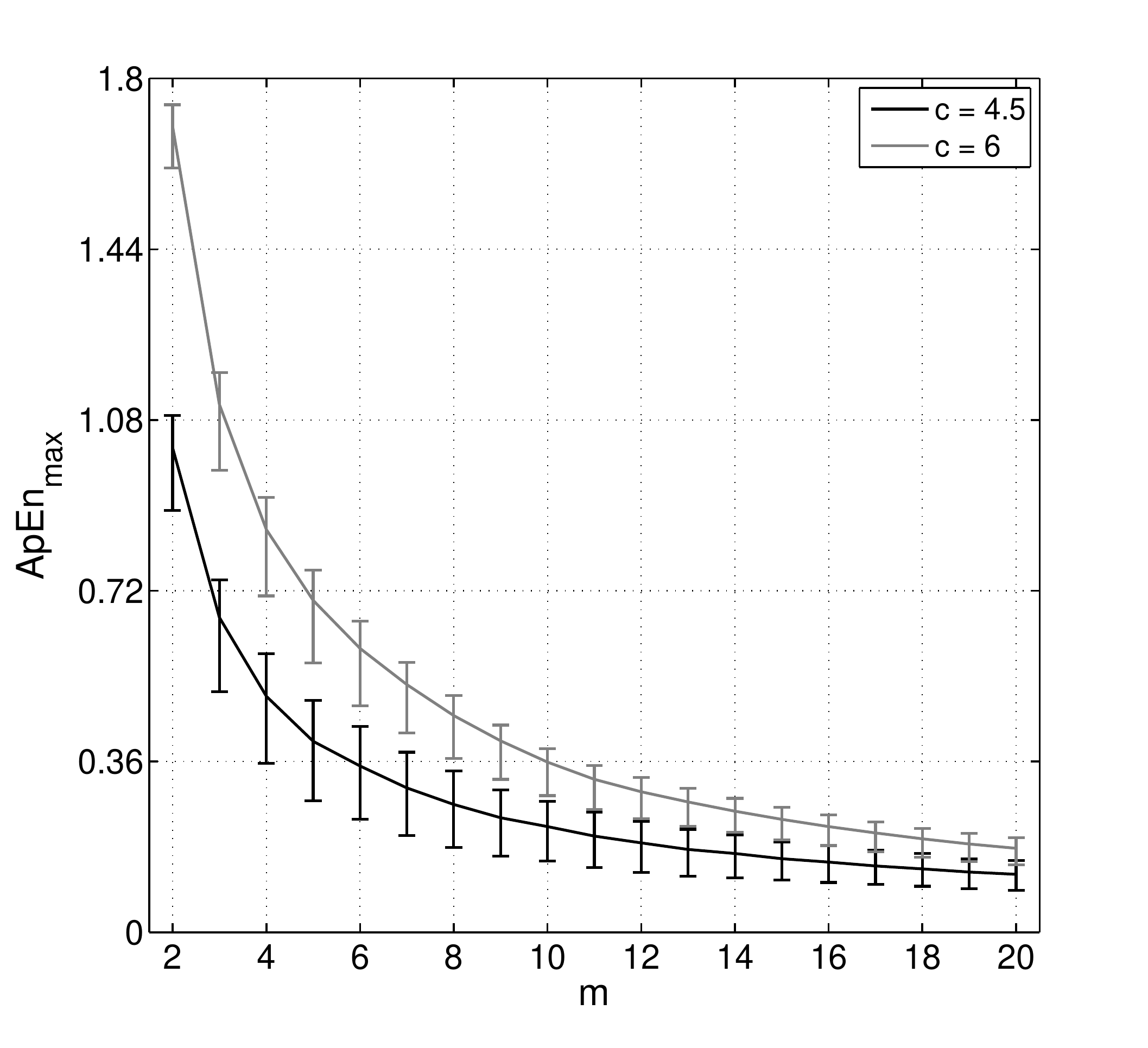}\label{fig:2.1}}\hspace{-5mm}
    \subfloat[][]{\includegraphics[width=0.500\linewidth,height=\textheight,keepaspectratio]{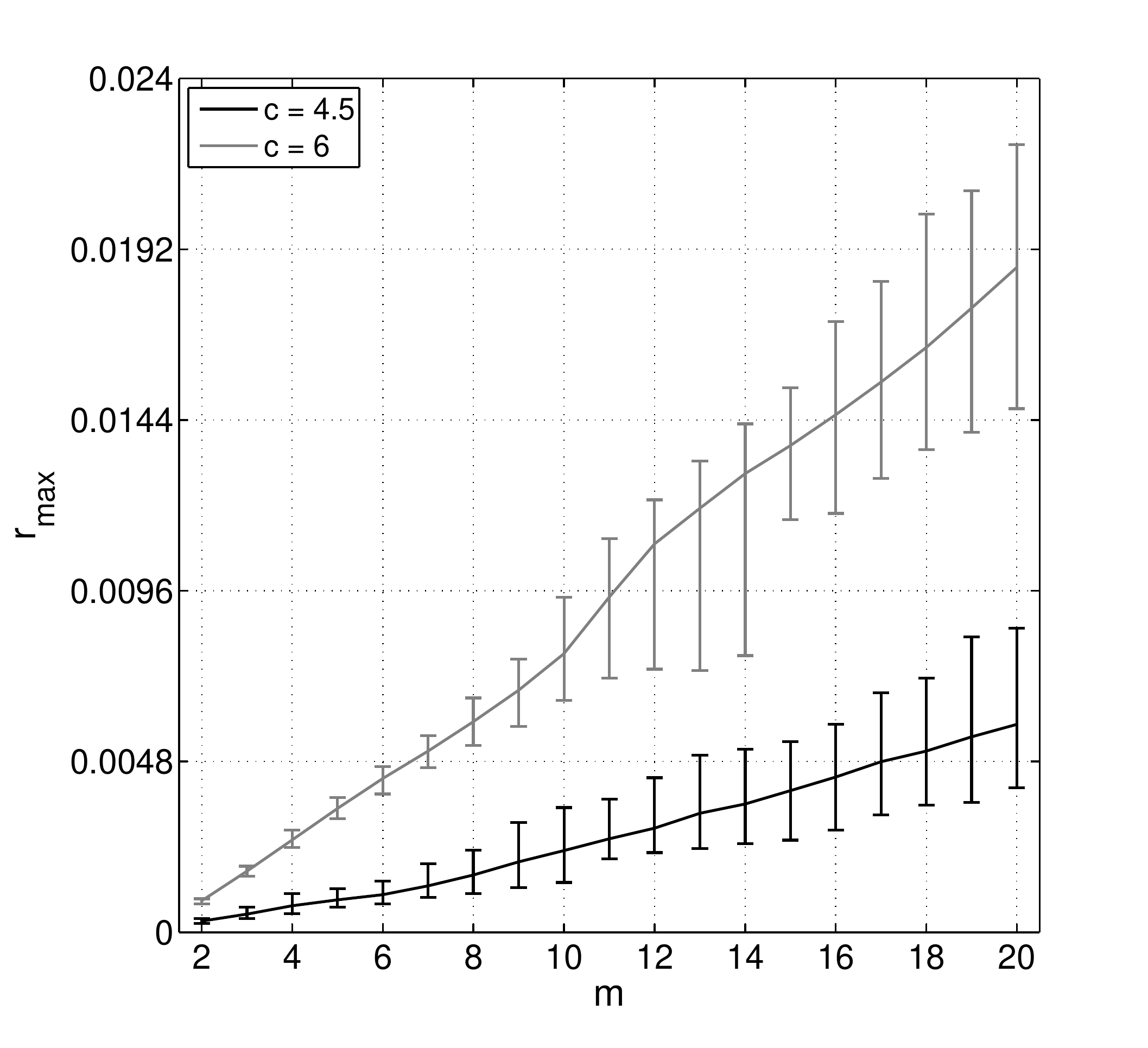}\label{fig:2.2}}
    \vspace{-2mm}
    \subfloat[][]{\includegraphics[width=0.500\linewidth,height=\textheight,keepaspectratio]{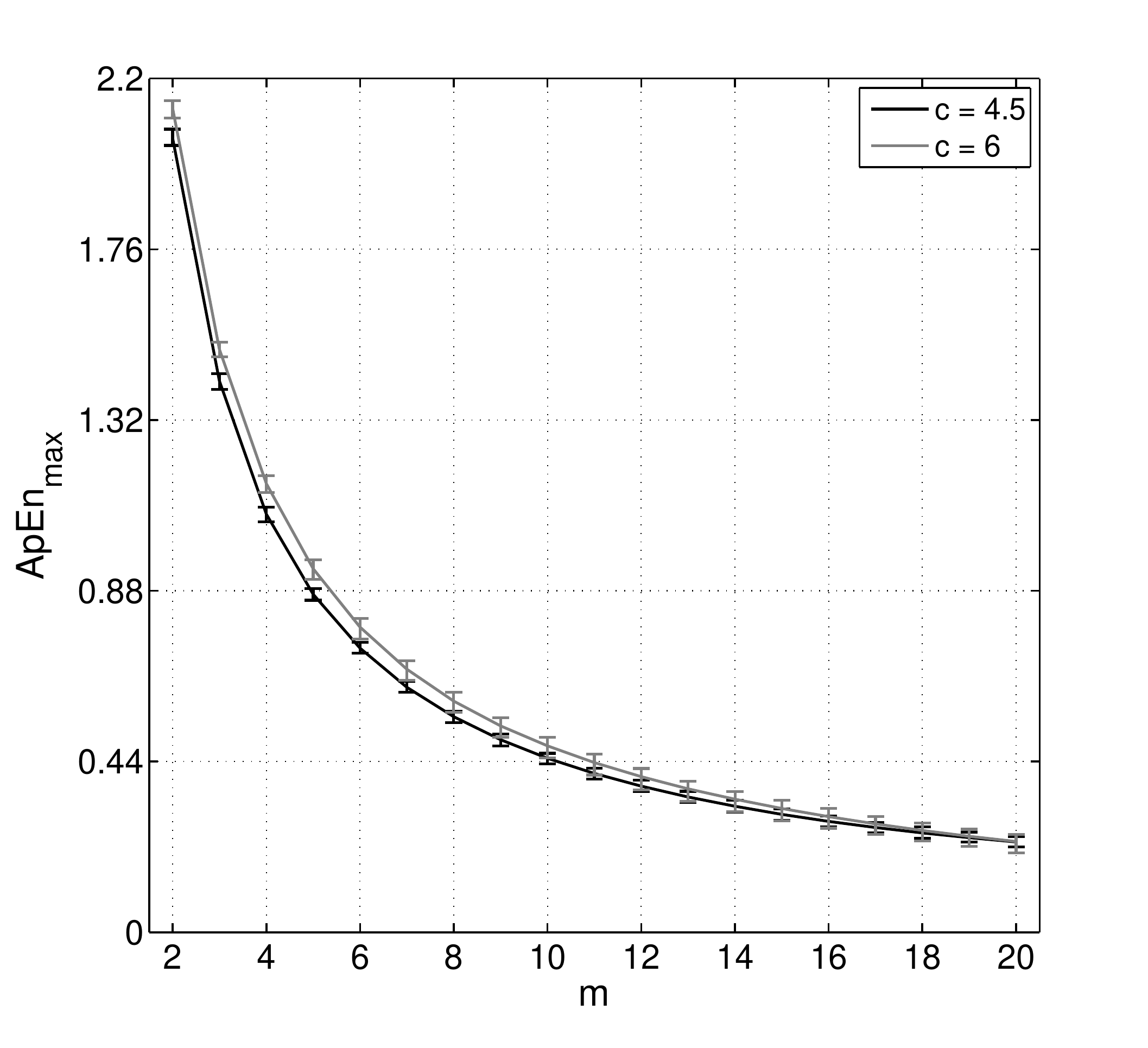}\label{fig:2.3}}\hspace{-5mm}
    \subfloat[][]{\includegraphics[width=0.500\linewidth,height=\textheight,keepaspectratio]{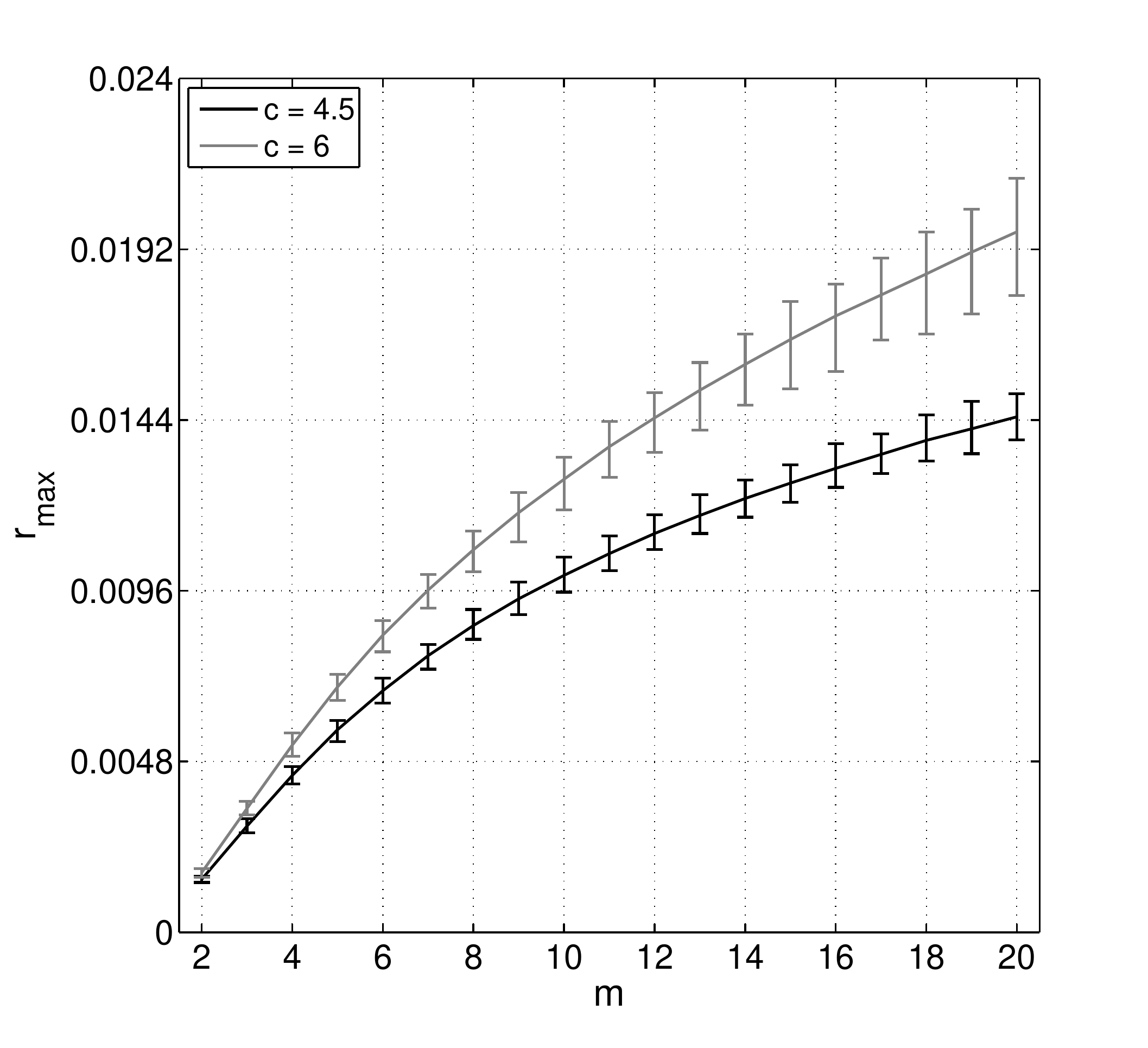}\label{fig:2.4}}
    \caption{Mackey-Glass model:  mean  and $95\%$  confidence interval.  Noiseless:  (a)  $ApEn_{max}$.  (b) $r_{max}$.
    With SNR$=5$~dB: (c) $ApEn_{max}$.  (d) $r_{max}$.}
    \label{fig:2}
\end{figure}

A real  physiological signal recorded  using stereo electroencephalography  (EEG) with eight  multilead electrodes (2~mm
long and  1.5~mm apart)  was studied.  It  was filtered and  amplified using  a 1-40  Hz band-pass  filter.  A four-pole
Butterworth filter was used as anti-aliasing low-pass filter.  This signal was digitized at 256 Hz through a 10 bits A/D
converter.  A physician accomplished  the analysis of pre-ictal and  ictal data by visual inspection  of the EEG record.
According to  the visual assessment of  the EEG seizure recording,  the  patient presented an epileptogenic  area in the
hippocampus with immediate propagation to the girus cingular  and the supplementary motor area,  on the left hemisphere.
In Fig.~\ref{fig:1},  the EEG signal of  two ictal and two pre-ictal episodes corresponding to  a depth electrode in the
hippocampus is presented.  All these  episodes  contains  3000  data  samples.  The  first  pre-ictal and ictal episodes
comprise the signal portions for $n\in[1500,4500]$ and $n\in[5000,8000]$ data points respectively.  The second pre-ictal
and ictal portions  were selected for $n\in[18000,21000]$  and $n\in[21400,24400]$ respectively.  Each of  the data sets
were normalized to have unitary energy and the $\tau$  parameter was selected as described above among the four signals.
$ApEn_{max}$ and $r_{max}$ were  then calculated for $2\leq m\leq20$.  Additionally,  white Gaussian  noise was added to
the raw EEG signal with \textit{SNR}$=5$ ~dB (the actual \textit{SNR} of the EEG signal is unknown) and $ApEn_{max}$ and
$r_{max}$ were calculated again.
\begin{figure}[!ht]
    \centering
    \subfloat[][]{\includegraphics[width=0.500\linewidth,height=\textheight,keepaspectratio]{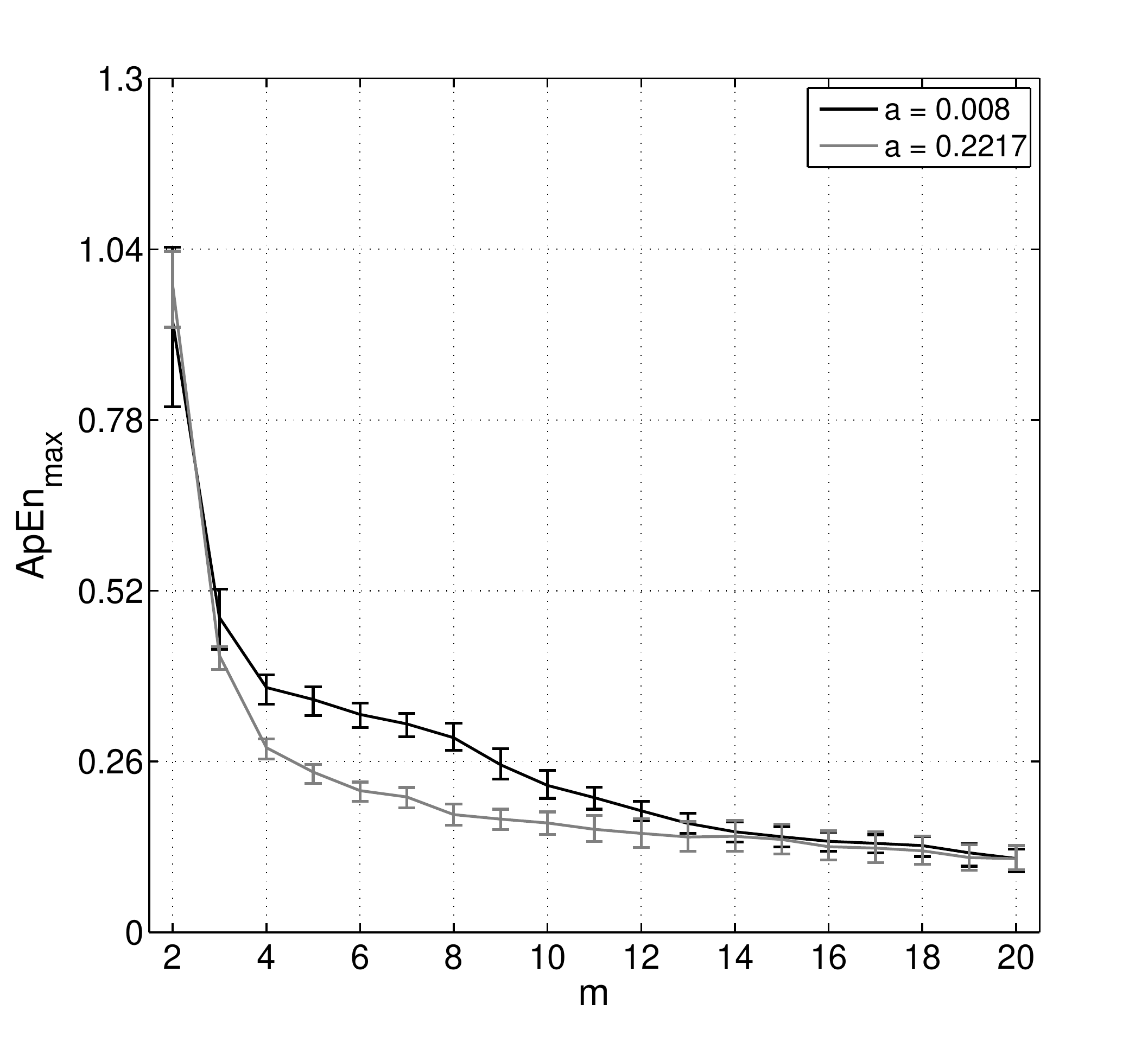}\label{fig:3.1}}\hspace{-5mm}
    \subfloat[][]{\includegraphics[width=0.500\linewidth,height=\textheight,keepaspectratio]{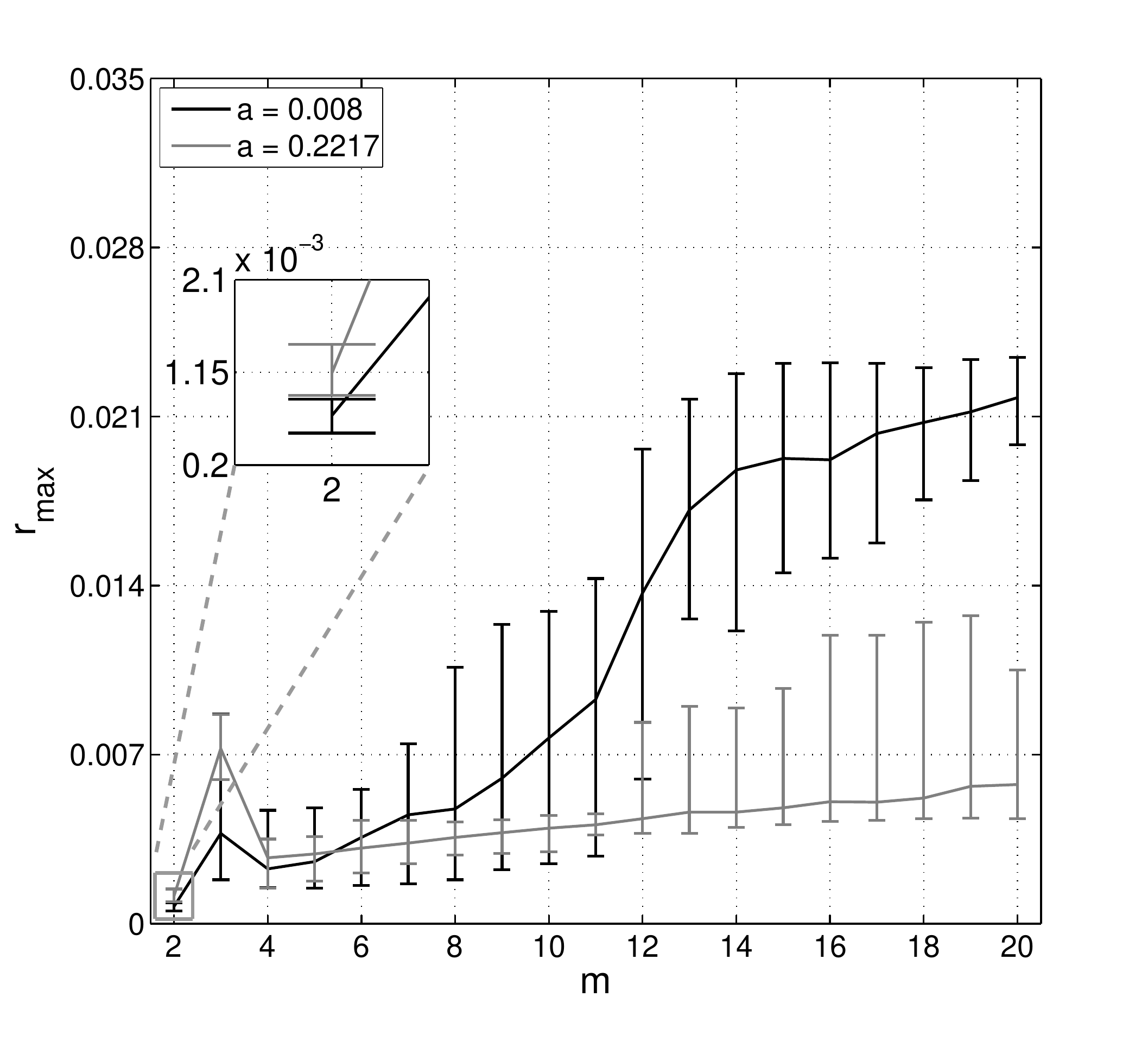}\label{fig:3.2}}
    \vspace{-2mm}
    \subfloat[][]{\includegraphics[width=0.500\linewidth,height=\textheight,keepaspectratio]{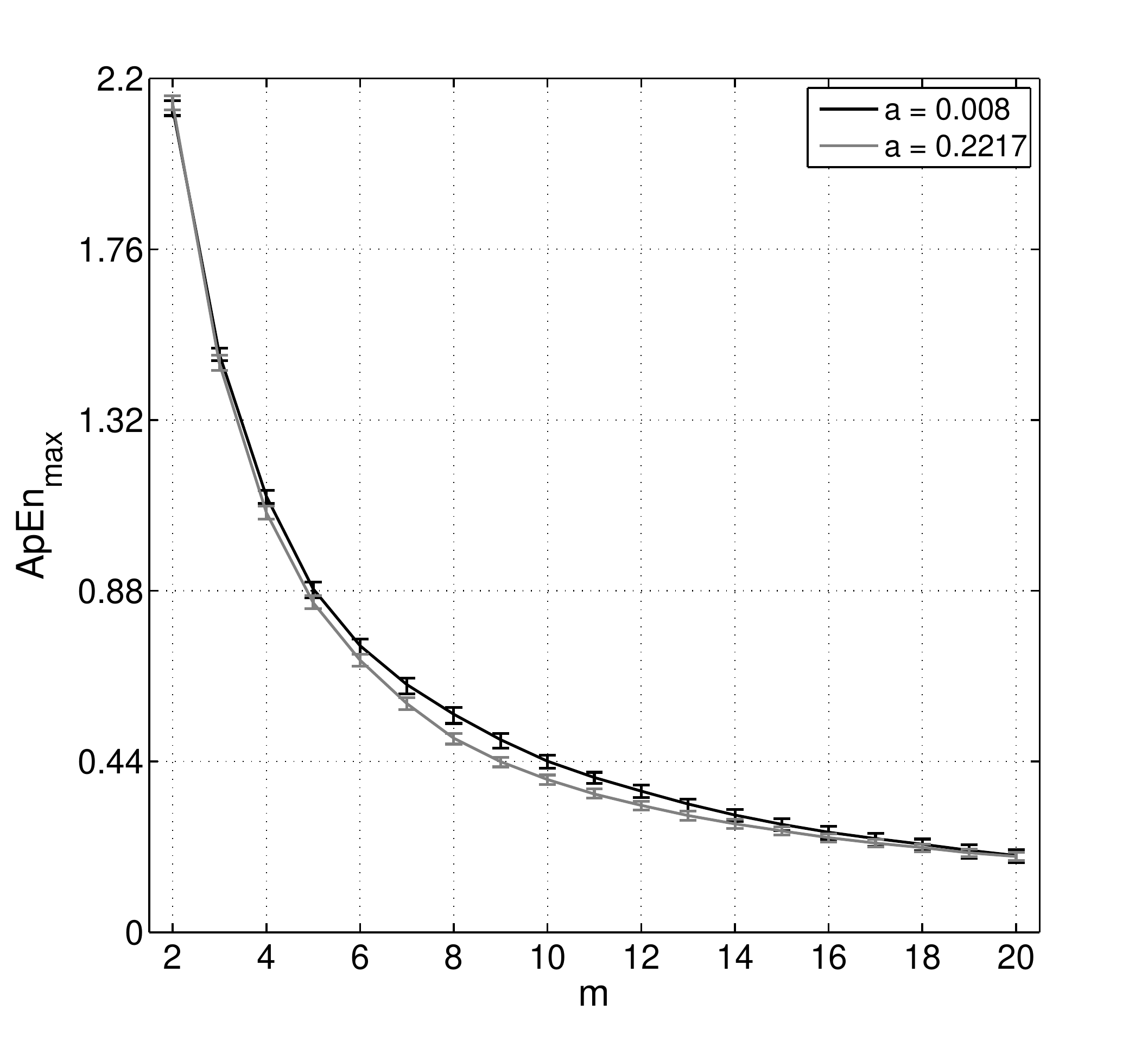}\label{fig:3.3}}\hspace{-5mm}
    \subfloat[][]{\includegraphics[width=0.500\linewidth,height=\textheight,keepaspectratio]{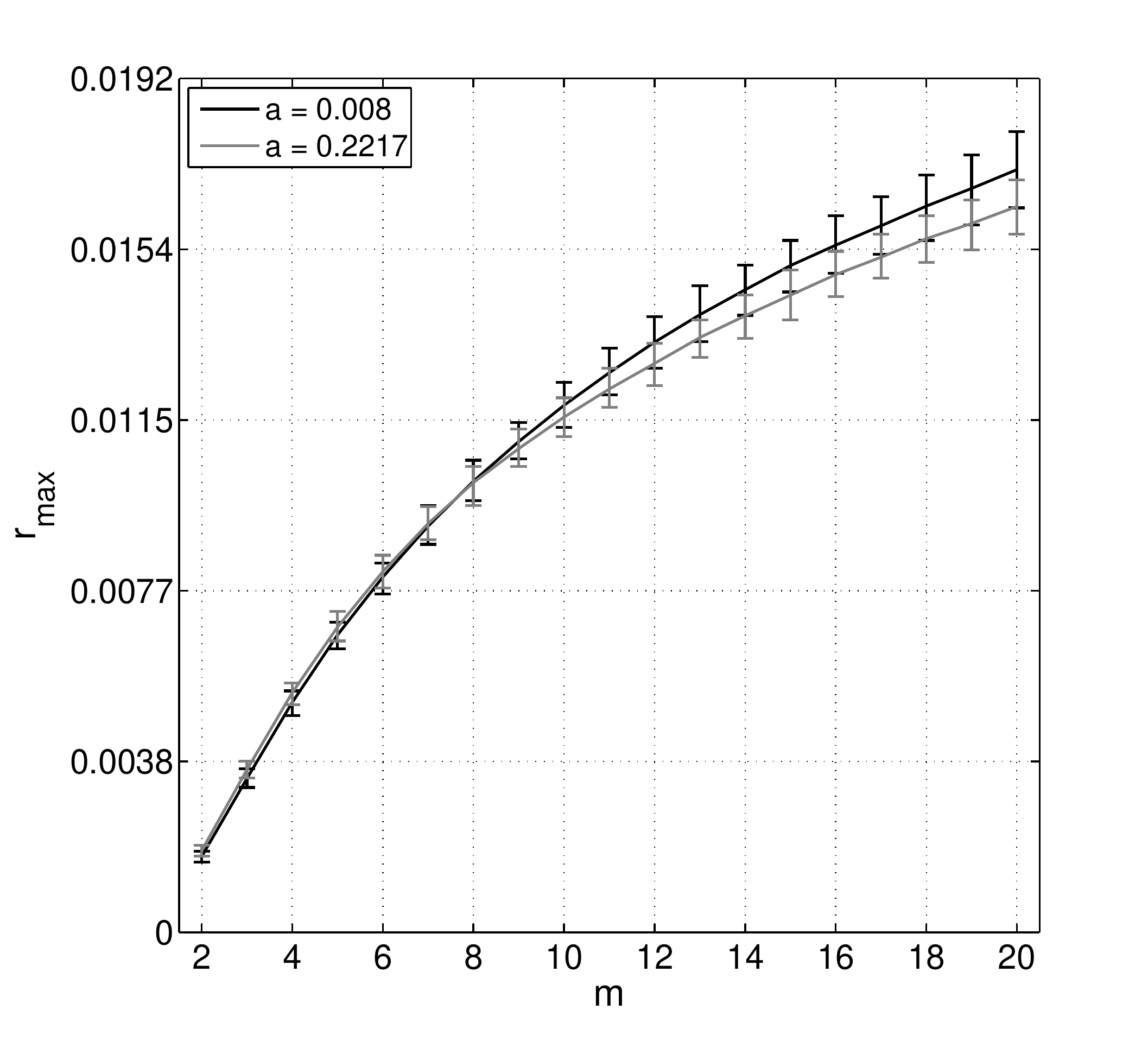}\label{fig:3.4}}
    \caption{Shilnikov's  type chaos  model:  mean and  $95\%$ confidence  interval.  Noiseless:  (a) $ApEn_{max}$.  (b)
    $r_{max}$.  With \textit{SNR}$=5$~dB:  (c) $ApEn_{max}$.  (d)  $r_{max}$.  In  (b)  an  enlarged  view  for $m=2$ is
    presented.}
    \label{fig:3}
\end{figure}
\section{Results and Discussion}
\label{sec:res}
\begin{figure}[!t]
    \centering
    \subfloat[][]{\includegraphics[width=0.500\linewidth,height=\textheight,keepaspectratio]{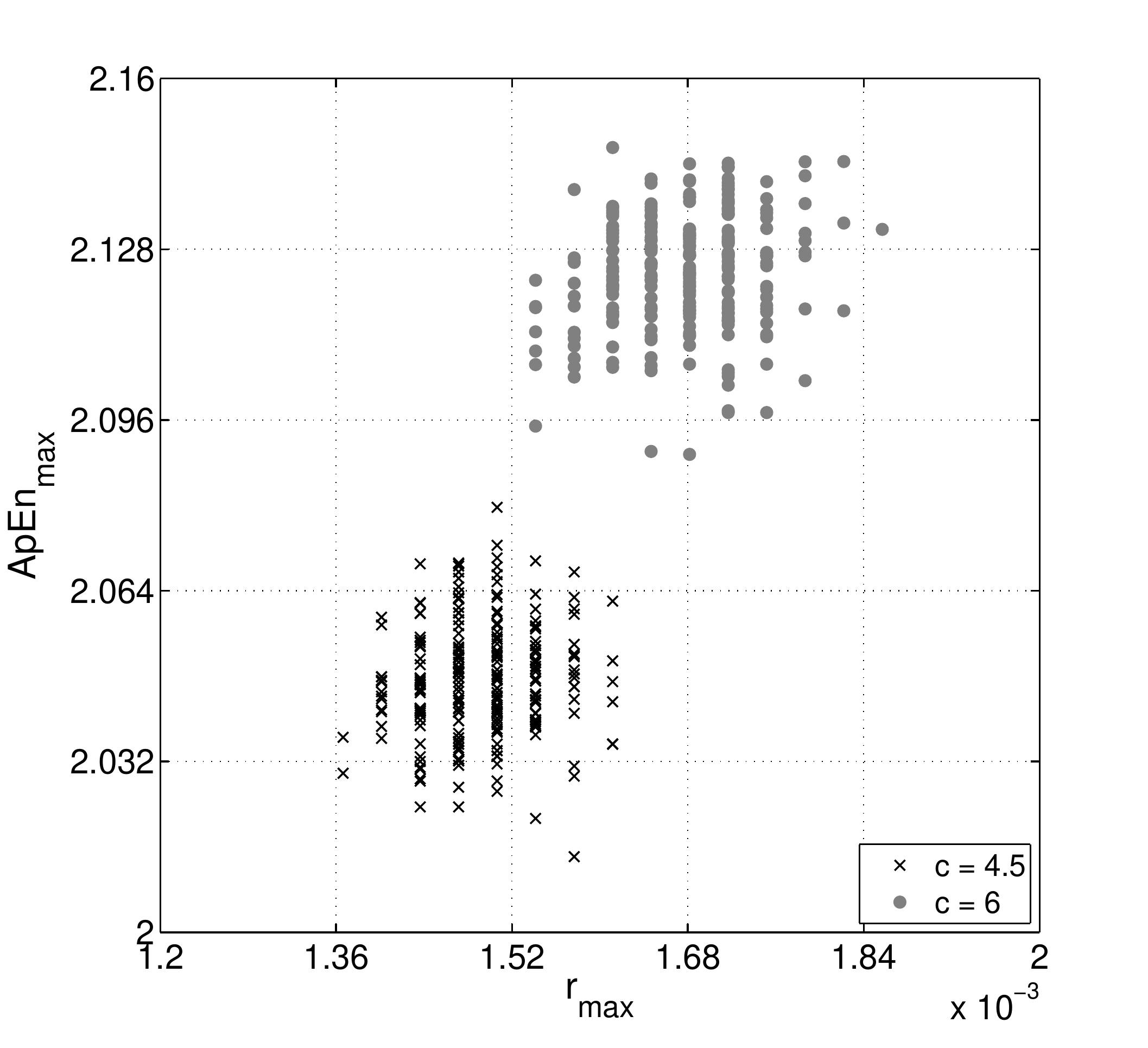}\label{fig:4.1}}\hspace{-5mm}
    \subfloat[][]{\includegraphics[width=0.500\linewidth,height=\textheight,keepaspectratio]{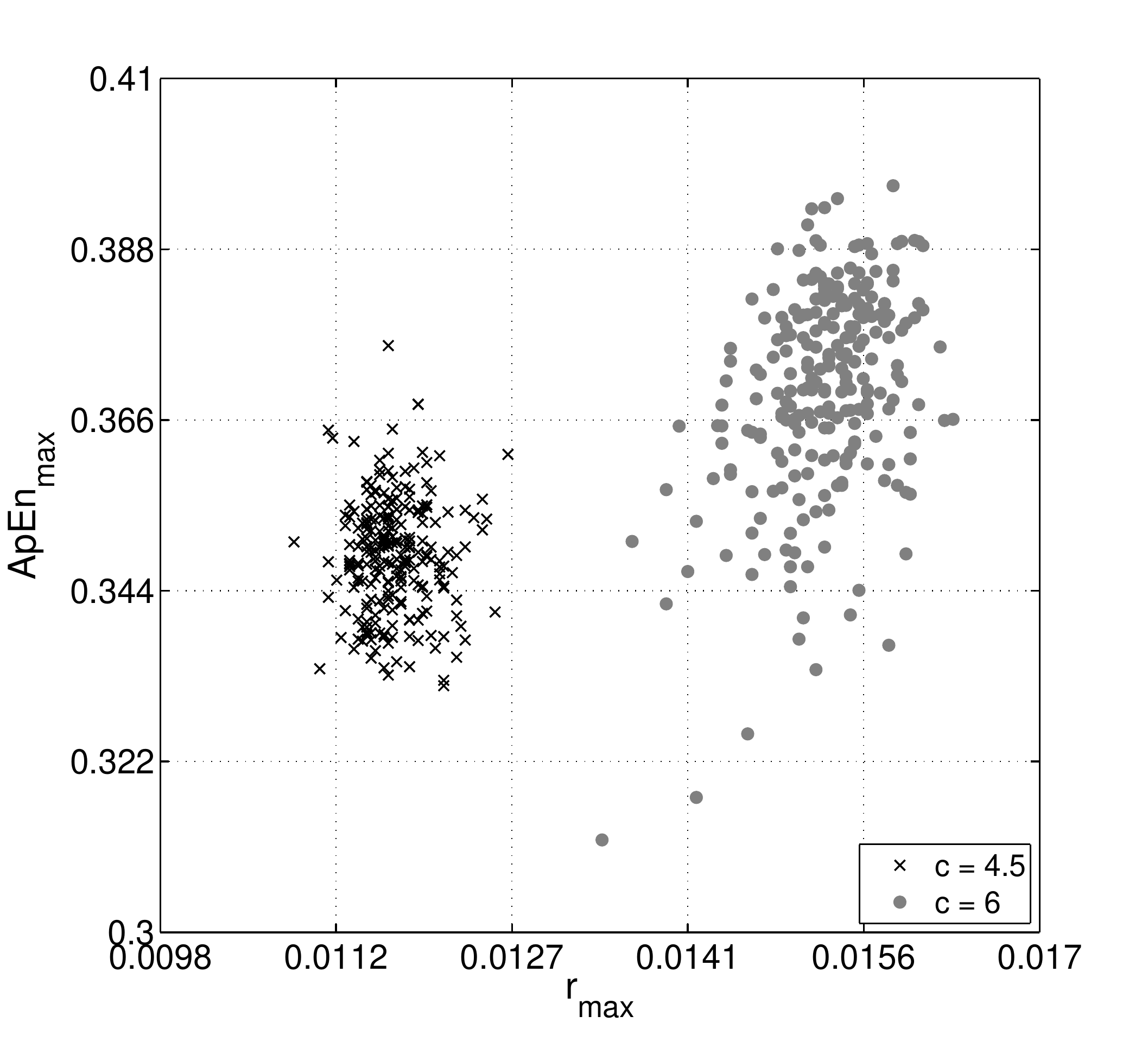}\label{fig:4.2}}
    \vspace{-2mm}
    \subfloat[][]{\includegraphics[width=0.500\linewidth,height=\textheight,keepaspectratio]{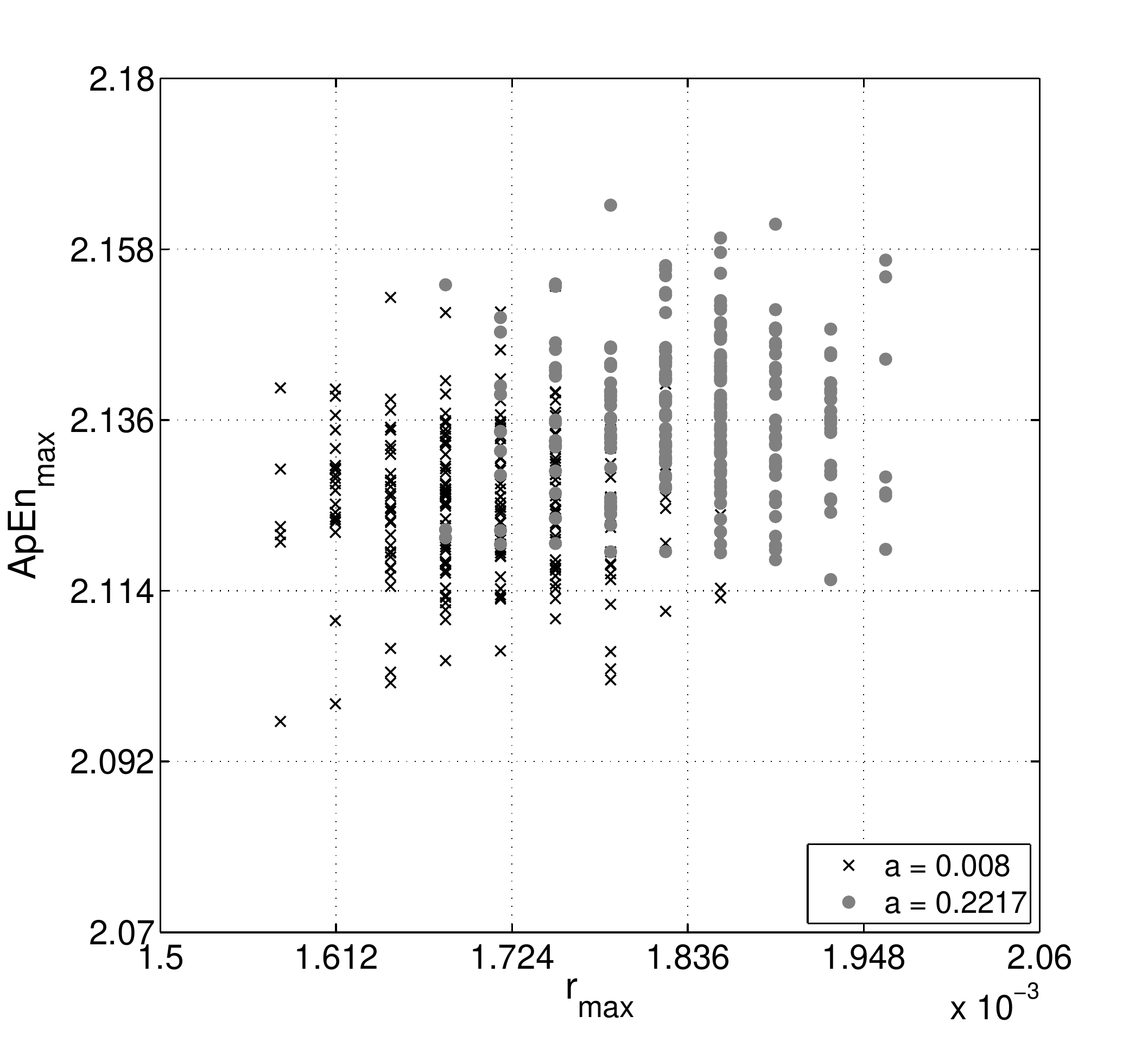}\label{fig:4.3}}\hspace{-5mm}
    \subfloat[][]{\includegraphics[width=0.500\linewidth,height=\textheight,keepaspectratio]{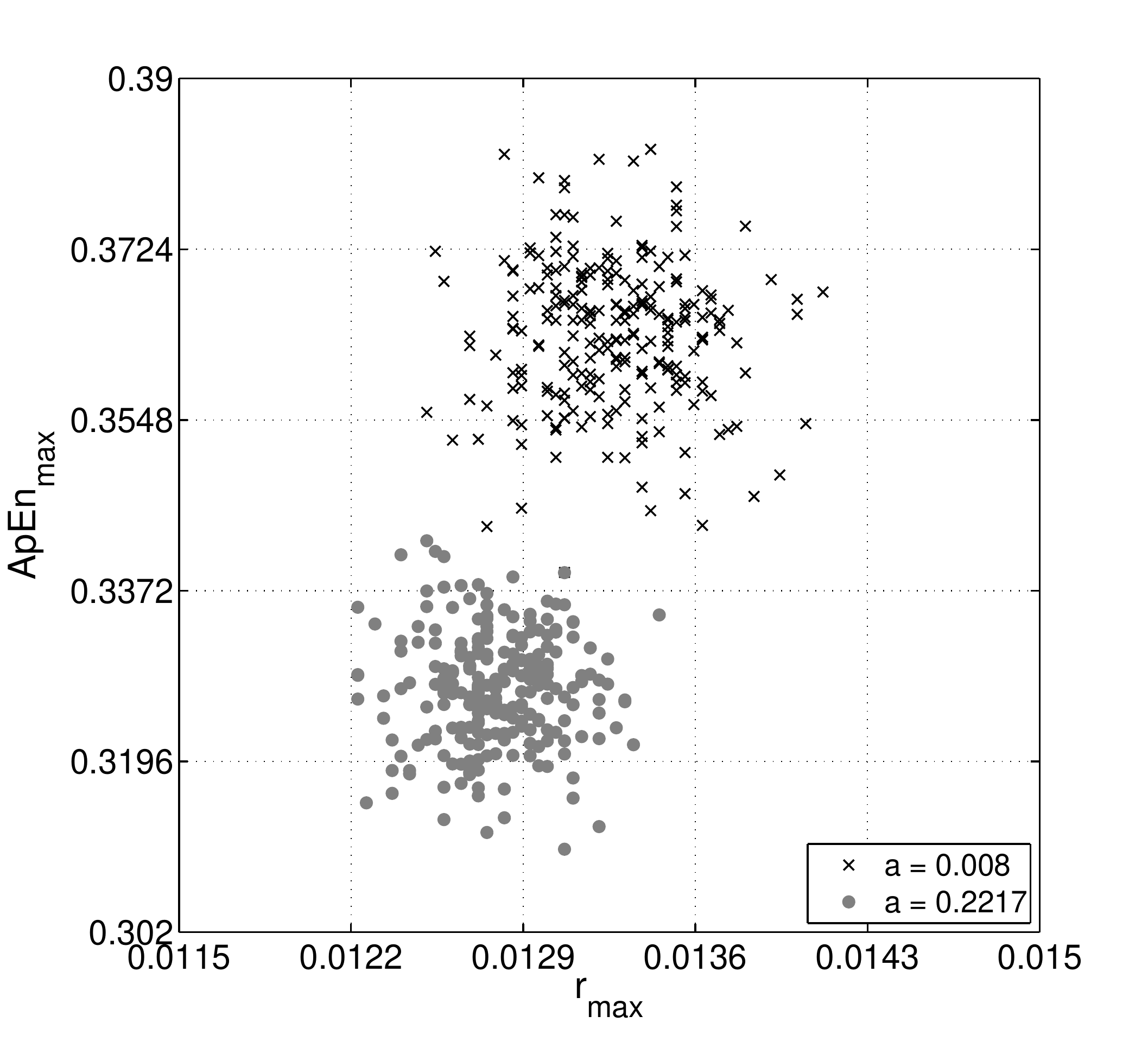}\label{fig:4.4}}
    \caption{  \textit{SNR}$=5$~dB,  $ApEn_{max}$  \textit{vs}  $r_{max}$  plot:  (a)  Mackey-Glass  model  $m=2$.   (b)
    Mackey-Glass model $m=12$.  (c) Shilnikov's type model $m=2$.  (d) Shilnikov's type model $m=12$.}
    \label{fig:4}
\end{figure}

Fig.~\ref{fig:2} summarizes the results obtained for the Mackey-Glass model simulations.  The $ApEn_{max}$ and $r_{max}$
mean and $95\%$ confidence interval (CI) are presented as  functions of $m$ for two different $c$ parameter values.  The
CIs were empirically  obtained by sorting the $ApEn_{max}$  and $r_{max}$ values calculated from  the $240$ realizations
and taking the $2.5\%$ and the $97.5\%$ quantiles as the lower and upper bound respectively.  In Fig.~\ref{fig:2.1},  it
can be noticed that the curves of $ApEn_{max}$ become closer as $m$ increases,  achieving the maximum distance at $m=2$.
On the contrary,  in Fig.~\ref{fig:2.2} it can be observed that the distance between the $r_{max}$ curves becomes larger
as $m$ increases.  Figs.~\ref{fig:2.3} and  \ref{fig:2.4}  show  the  effect  of  noise  over $ApEn_{max}$ and $r_{max}$
estimations.  First,  notice that,  compared  to noise free figures,  the  mean values of $ApE_{max}$  and $r_{max}$ are
increased due to the addition of noise.  Additionally,  in both cases,  the CIs are reduced.  Finally,  the $ApEn_{max}$
and $r_{max}$ curves for different $c$ parameter values are closer  to each other than in the case without noise.  It is
important  to remark  that,  in presence  of noise,  while  $r_{max}$ is  still able  to discern  between dynamics,  the
discrimination capacity  of $ApEn_{max}$ is  highly reduced.  In conclusion,  these  results suggest that  $r_{max}$ can
bring useful information even in presence of noise.

Shilnikov's type chaos model results are presented  in Fig.~\ref{fig:3}.  In Fig.~\ref{fig:3.1},  it can be noticed that
it is impossible to distinguish the two dynamics using $ApEn_{max}$ calculated with $m=2$.  Nevertheless,  embedding the
system in a  higher dimension such as $m=3$  (minimum embedding dimension),  distinctions between dynamics  can be made.
However,  in Fig.~\ref{fig:3.2},  it can be seen that for $m=2$, $r_{max}$ indicates a difference between dynamics.  The
added noise  has the same above  mentioned influence over both  $ApEn_{max}$ and $r_{max}$  (see Figs.~\ref{fig:3.3} and
\ref{fig:3.4}).  However,  in this  case the $r_{max}$  curves are closer  than those corresponding  to the Mackey-Glass
system.  From this simulation we can conclude that using $ApEn_{max}$ or $r_{max}$ independently can be inconvenient for
classification purposes.  Instead, we propose to study the combined use of both estimators for this task.

In  order to  illustrate the  advantages of  this new  approach,  Fig.~\ref{fig:4} shows  scatter plots  of $ApEn_{max}$
\textit{vs} $r_{max}$  for both models  with noise (\textit{SNR}$=5$~dB),  using  $m=2$ and $m=12$.  In  the presence of
noise,  it is  enough to set  $m=2$ and to use  only $ApEn_{max}$ to correctly  differentiate the two  dynamics from the
Mackey-Glass model (see Fig.~\ref{fig:2.3}).  However,  in Fig.~\ref{fig:4.1} it  can be noticed that $r_{max}$ provides
additional  information  that can  make  easier the  classification  process.  A  slightly  different  situation  can be
appreciated for Shilnikov's type dynamics.  Fig.~\ref{fig:4.3} shows that  it is not possible to discern between classes
using $ApEn_{max}$ calculated with $m=2$.  Nevertheless,  with the information brought by $r_{max}$, the two classes can
be separated  in a  more suitable  way.  As presented before,  when  there is  noise in  the signal,  the  assessment of
$ApEn_{max}$ using an $m$ value  equal or larger than the minimum embedding dimension could  be more accurate and robust
than just  setting $m=2$.  Cao's  algorithm suggests  that the  minimum embedding  dimension for  both models  should be
$m\approx 12$.  Such large $m$ value  is  the  result  of  noise  influence  in  the  estimation of the systems' minimum
embedding dimension.  The issue is that, for the Mackey-Glass model, $ApEn_{max}$ losses its discrimination capacity for
high $m$ values.  Nonetheless,  as can be appreciated in  Fig.~\ref{fig:4.2},  the two classes can be still successfully
separated using only $r_{max}$.  On the other hand, Fig.~\ref{fig:4.4} shows that for $m=12$, the two different dynamics
from the  Shilnikov's type  model can  be more conveniently  clustered using  $ApEn_{max}$ than  using $r_{max}$.  These
results remark the importance of using both estimators together instead of each one individually.

\begin{figure}[!t]
    \hspace{-1.6cm}{
    \subfloat[][]{\includegraphics[width=0.390\linewidth,height=\textheight,keepaspectratio]{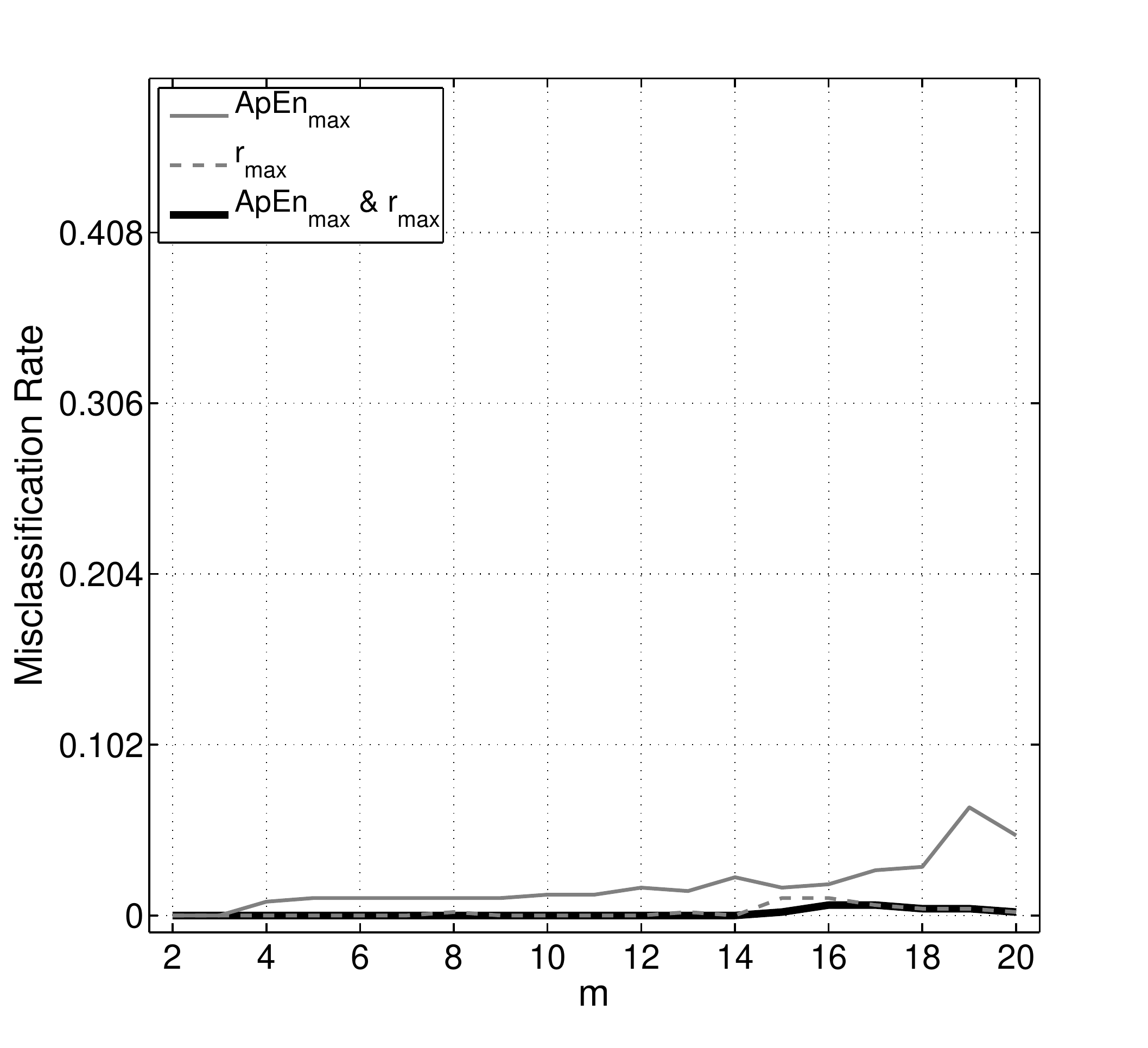}\label{fig:5.1}}\hspace{-5mm}
    \subfloat[][]{\includegraphics[width=0.390\linewidth,height=\textheight,keepaspectratio]{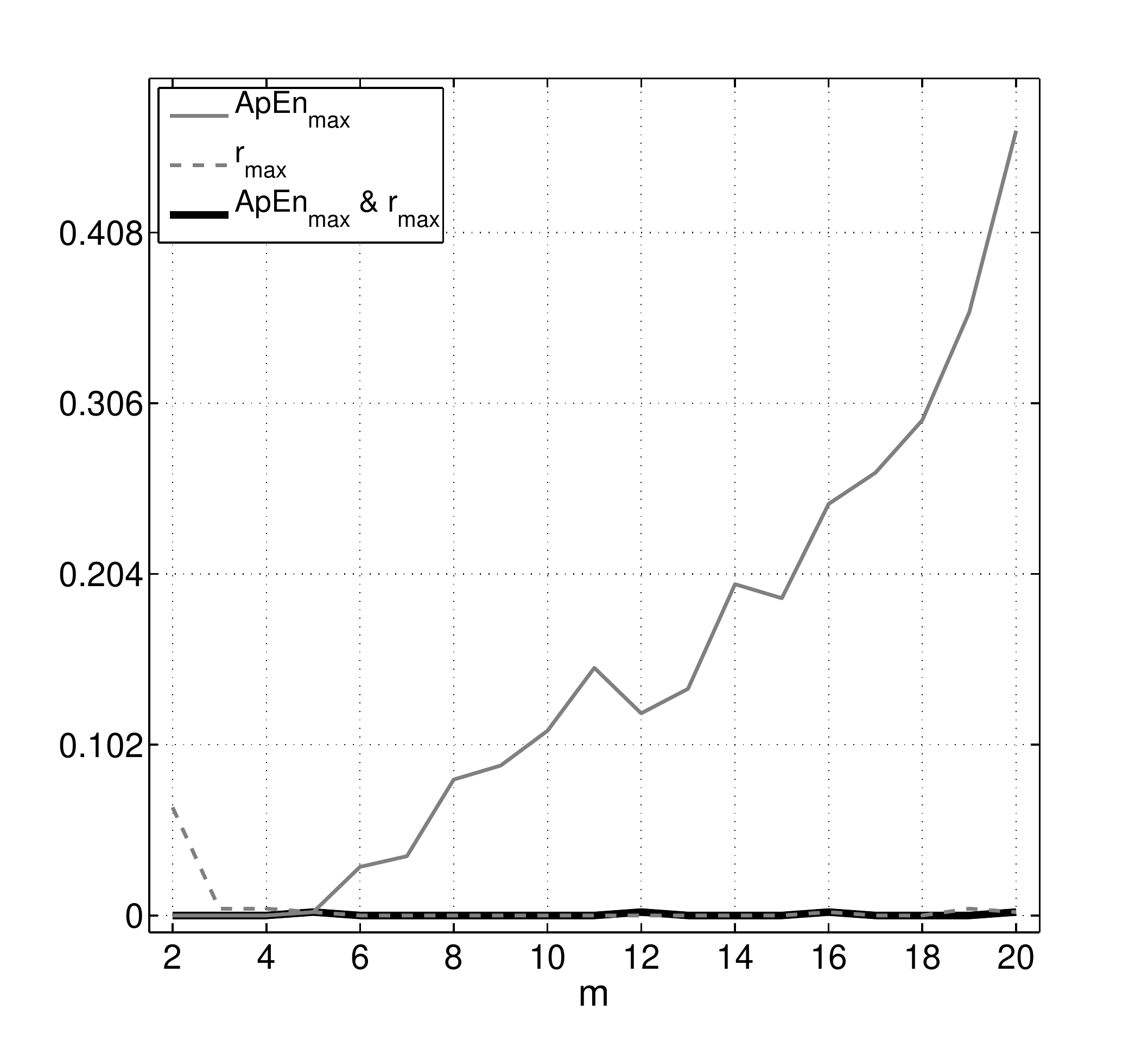}\label{fig:5.2}}\hspace{-5mm}
    \subfloat[][]{\includegraphics[width=0.390\linewidth,height=\textheight,keepaspectratio]{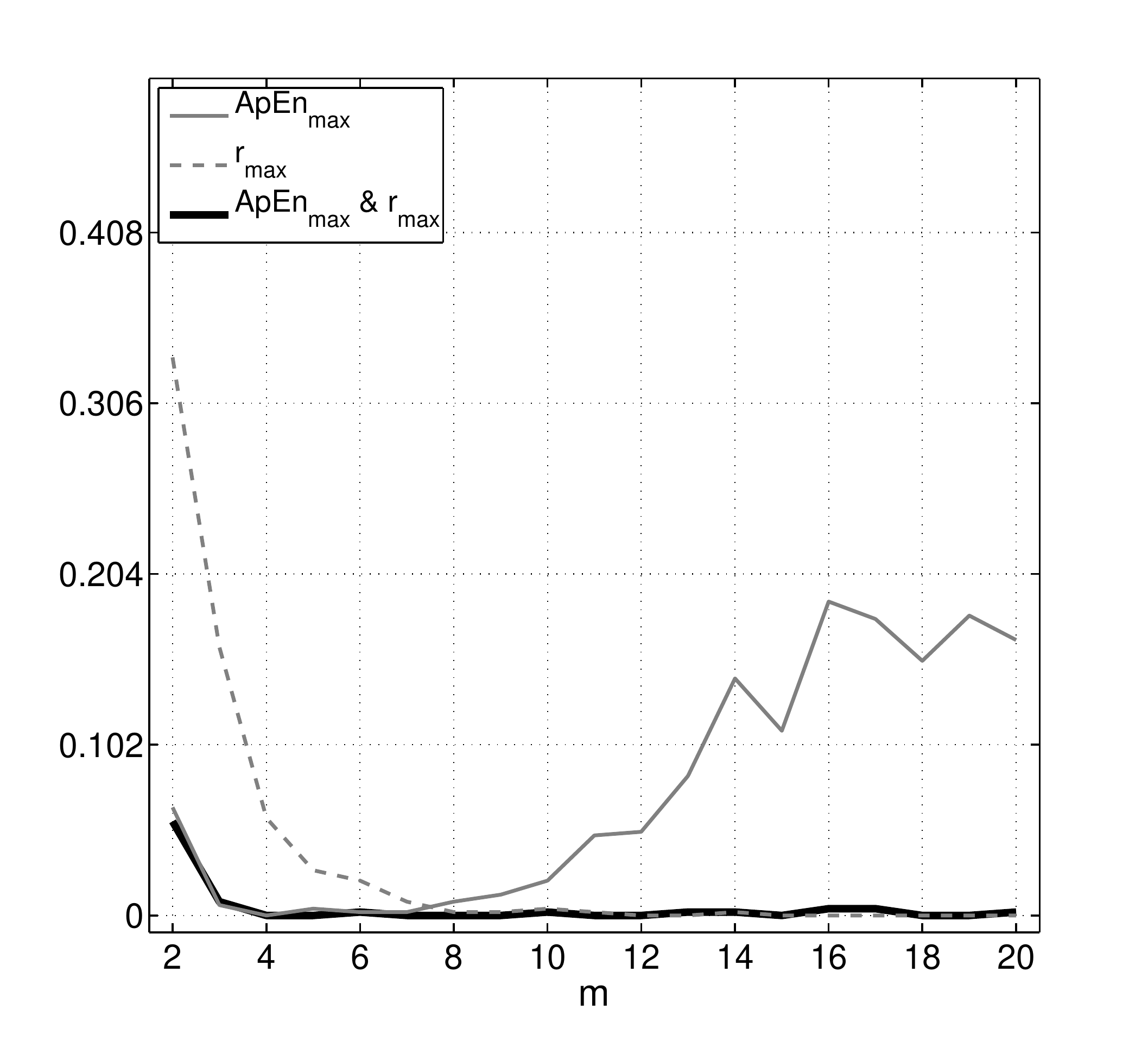}\label{fig:5.3}}\\}
    \vspace{-0.1mm}
    \hspace{-1.6cm}{
    \subfloat[][]{\includegraphics[width=0.390\linewidth,height=\textheight,keepaspectratio]{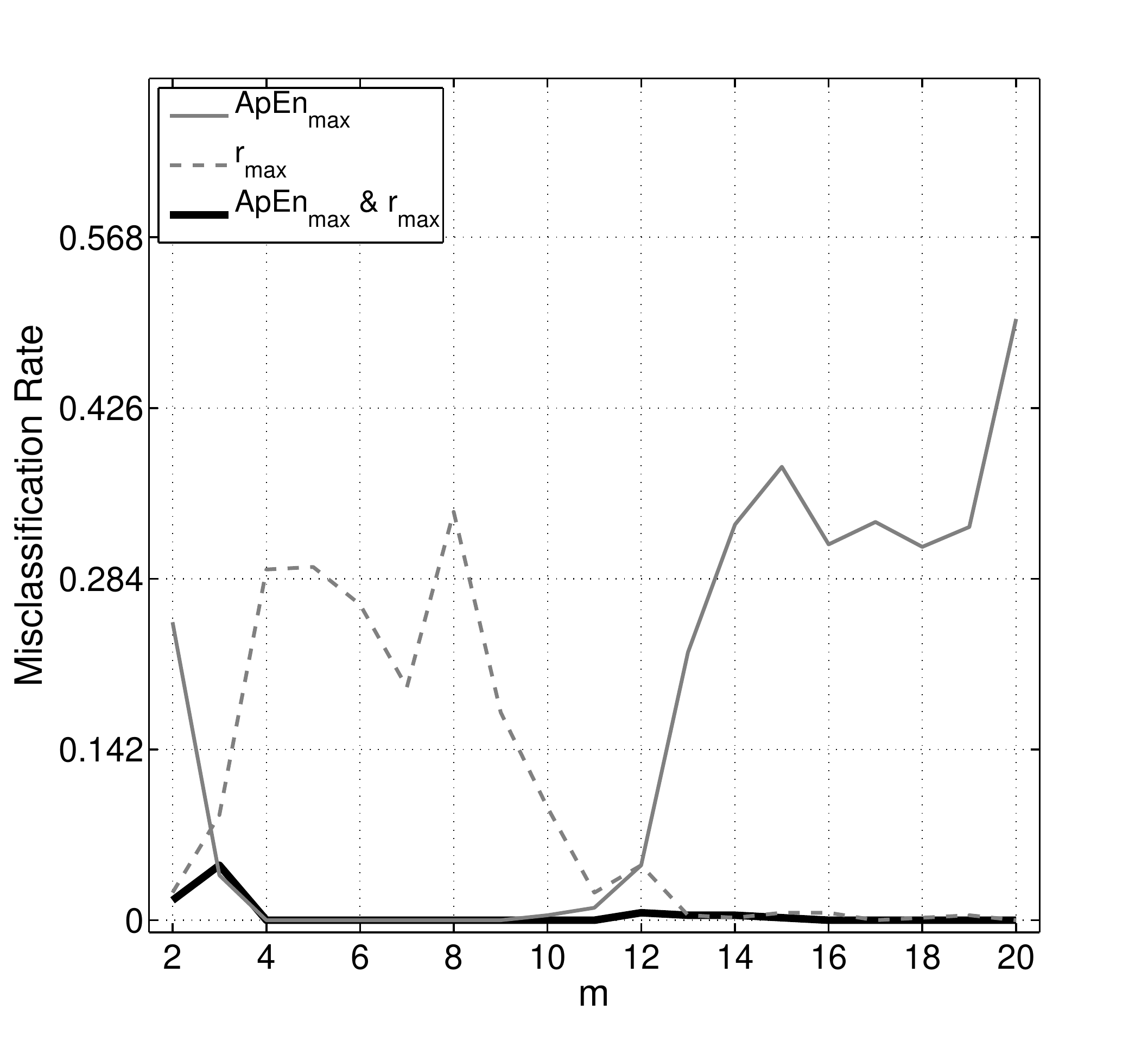}\label{fig:5.4}}\hspace{-5mm}
    \subfloat[][]{\includegraphics[width=0.390\linewidth,height=\textheight,keepaspectratio]{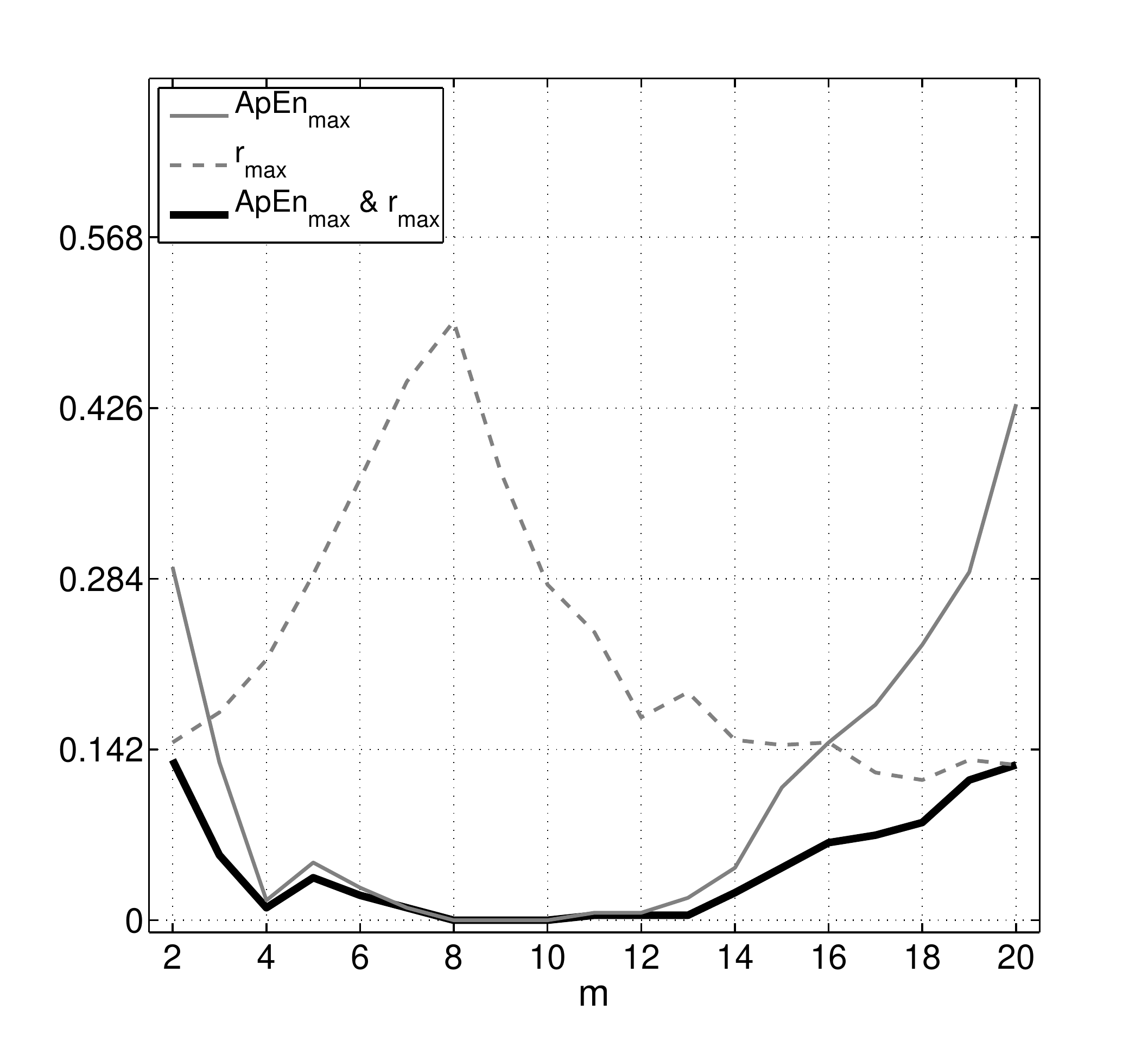}\label{fig:5.5}}\hspace{-5mm}
    \subfloat[][]{\includegraphics[width=0.390\linewidth,height=\textheight,keepaspectratio]{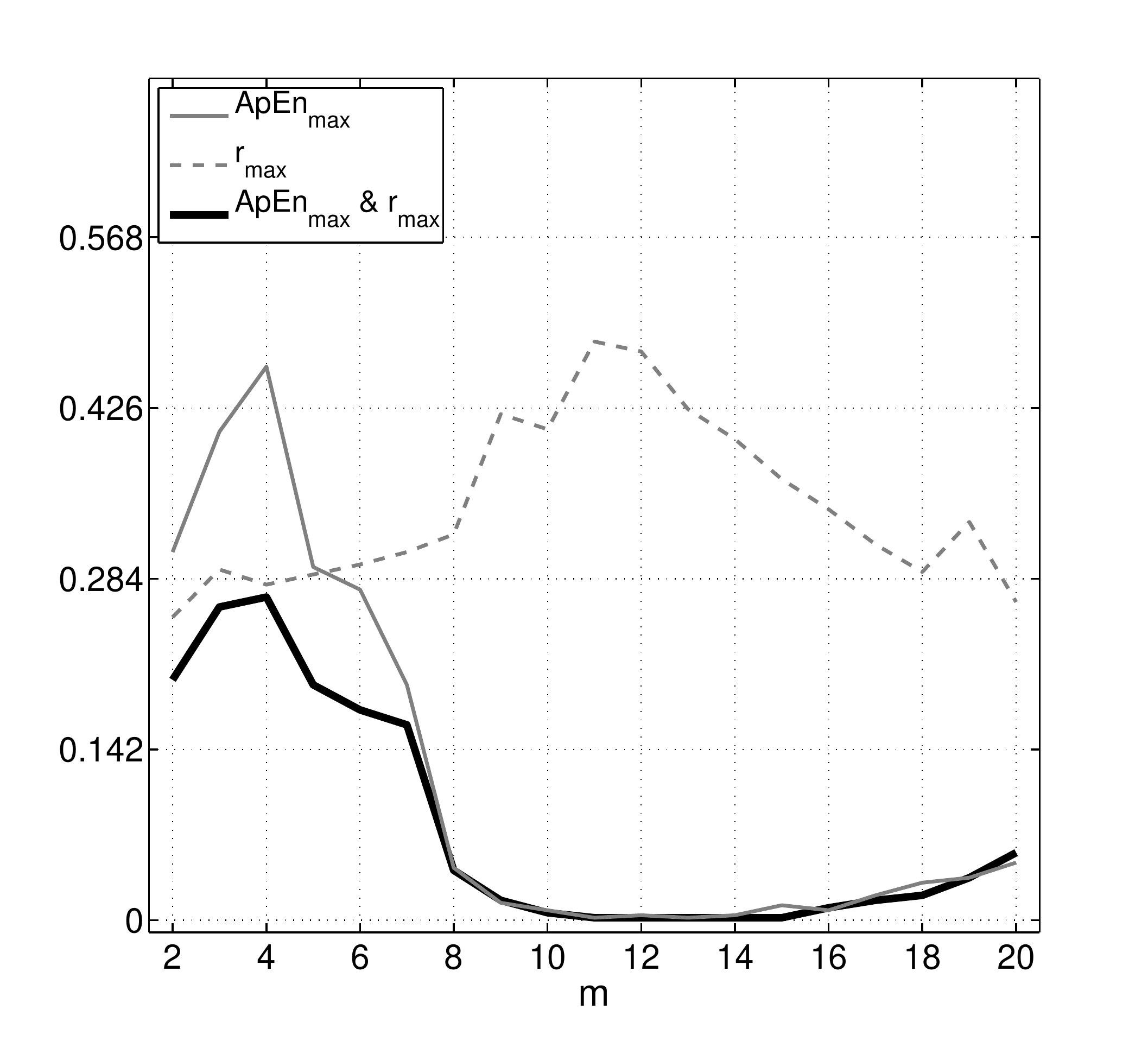}\label{fig:5.6}}\\}
    \vspace{-0.1mm}
    \hspace{-1.6cm}{
    \subfloat[][]{\includegraphics[width=0.390\linewidth,height=\textheight,keepaspectratio]{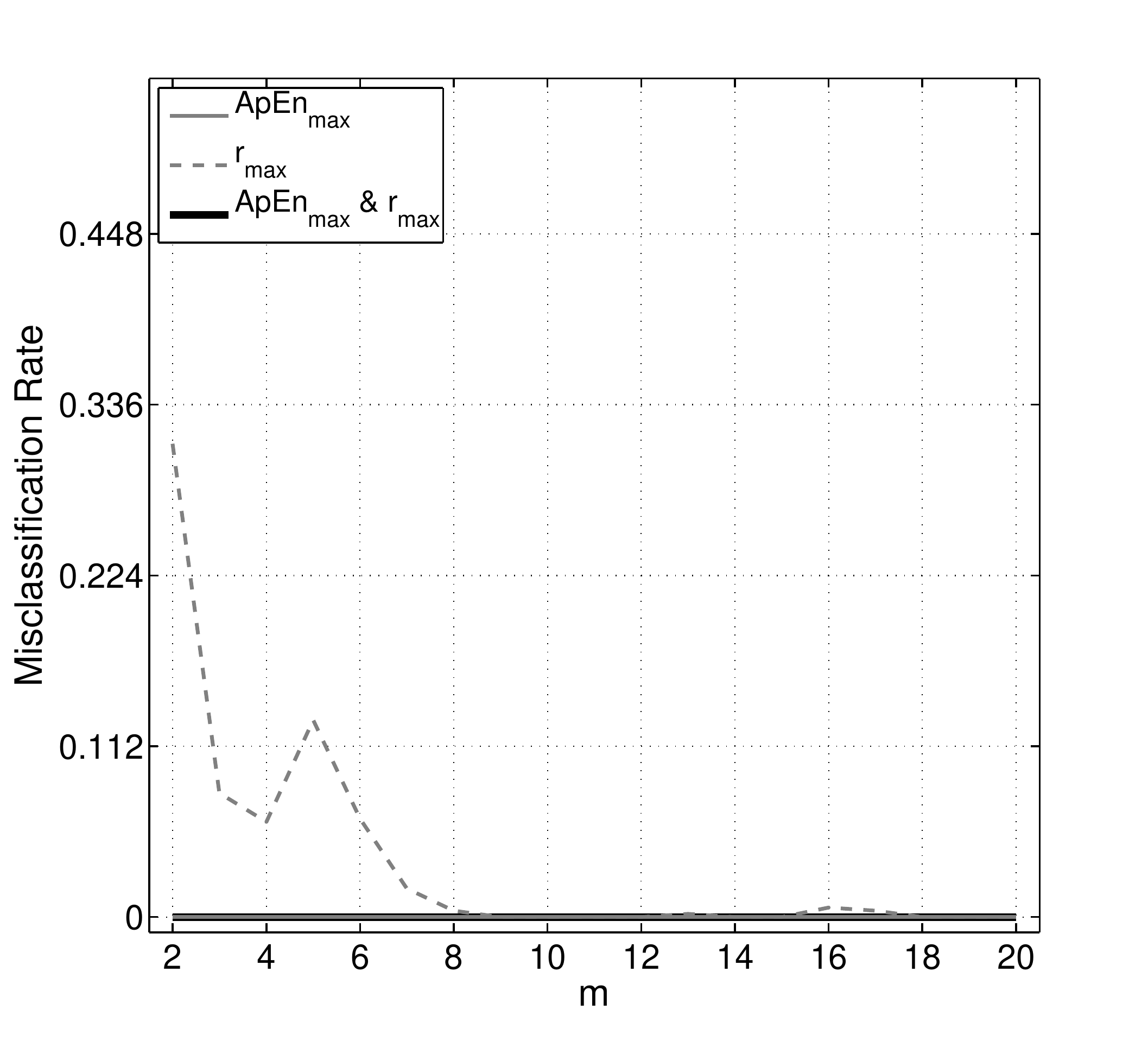}\label{fig:5.7}}\hspace{-5mm}
    \subfloat[][]{\includegraphics[width=0.390\linewidth,height=\textheight,keepaspectratio]{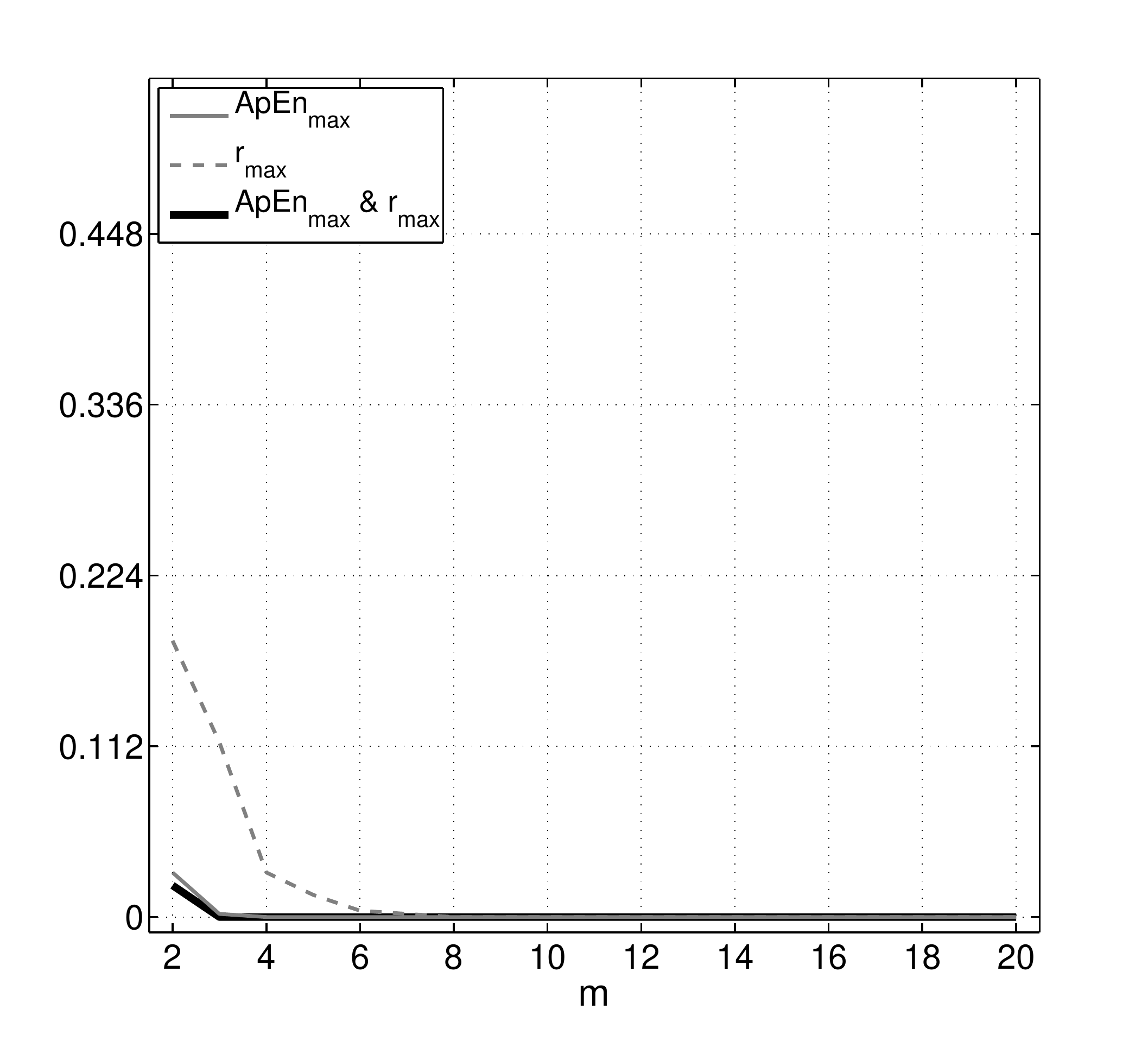}\label{fig:5.8}}\hspace{-5mm}
    \subfloat[][]{\includegraphics[width=0.390\linewidth,height=\textheight,keepaspectratio]{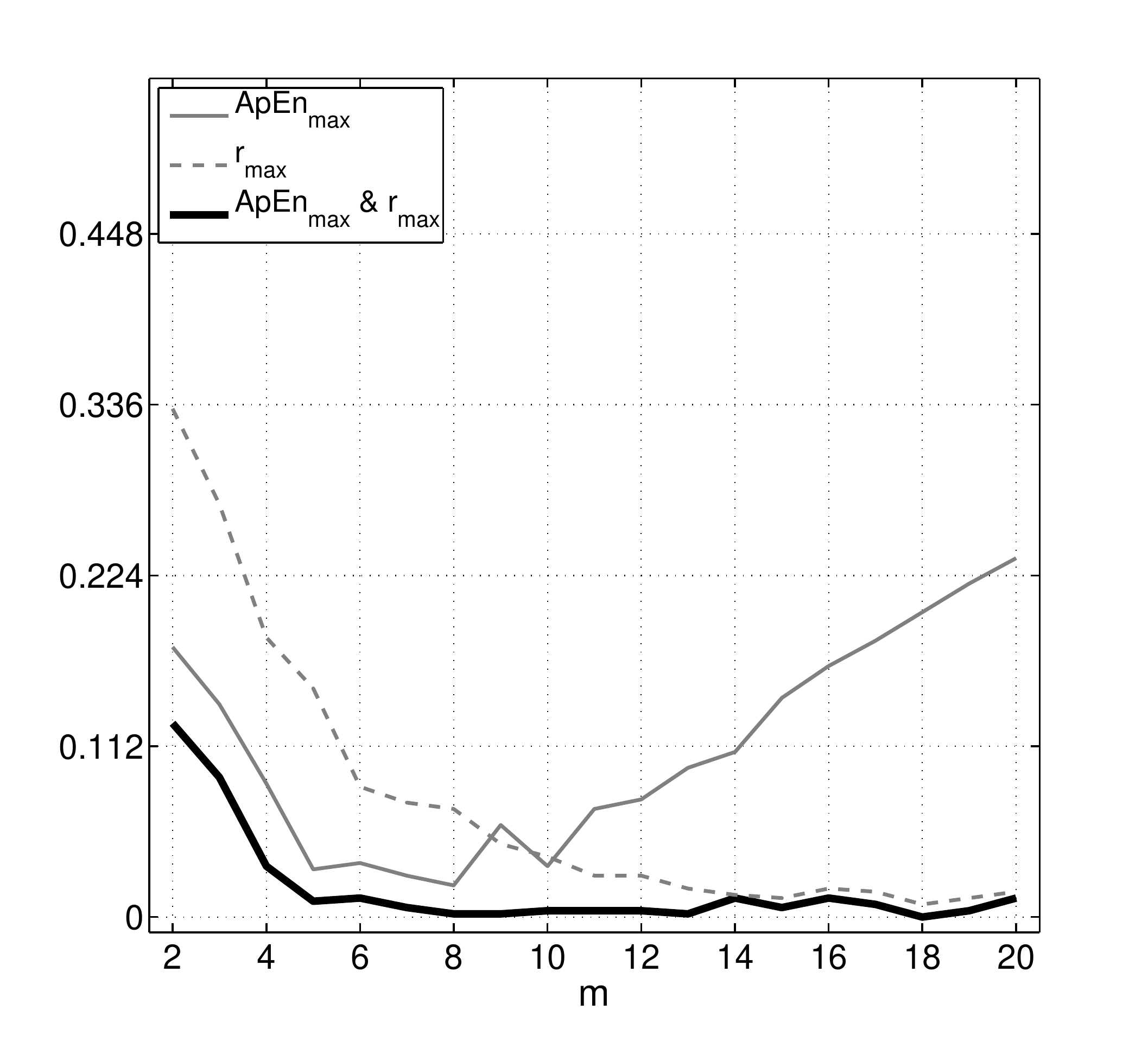}\label{fig:5.9}}\\}
    \caption{10-fold cross-validation  misclassification rate with a  linear \textit{SVM} classifier.  Mackey-Glass
    model:  (a)~noiseless.  (b)~\textit{SNR}$=5$~dB.  (c)~\textit{SNR}$=0$~dB.  Shilnikov's  type model:  (d)~noiseless.
    (e)~\textit{SNR}$=5$~dB.    (f)~\textit{SNR}$=0$~dB.    Logistic   map:   (g)~noiseless.    (h)~\textit{SNR}$=5$~dB.
    (i)~\textit{SNR}$=0$~dB.}
    \label{fig:5}
\end{figure}

In addition to  the Mackey-Glass and Shilnikov's systems  we introduce calculations of $ApEn_{max}$  and $r_{max}$ using
dynamics from the Logistic map,  $x_{n+1}=Rx_{n}(1-x_n)$,  in two chaotic regimes ($R=3.75$,  $R=3.95$) with and without
white Gaussian noise.  $ApEn_{max}$  and $r_{max}$ were evaluated  using the procedure aforementioned for  $2\leq m \leq
20$ with $\tau=1$ and $N=5000$.

With the goal of quantitatively verifying the  proficiency of $ApEn_{max}$ and $r_{max}$ as classification features,  we
perform a 10-fold cross-validation using linear \textit{support  vector machines} (SVMs).  We choose a linear classifier
given that its simplicity will disclose the real quality of the features.  The basic idea behind the SVMs is to separate
the classes using the  optimal hyperplane (the linear decision function that maximizes  the distance between the closest
points of different classes to the hyperplane) \cite{Vap2000}.  In Fig.~\ref{fig:5} the \textit{Misclassification Rates}
(MR)  for  three classifiers  as  functions of  $m$  and  different  noise  levels  (noiseless,  \textit{SNR}$=5$~dB and
\textit{SNR}$=0$~dB) are presented.  The first classifier uses only  $ApEn_{max}$ as input feature,  the second one uses
only $r_{max}$, and the third one uses jointly both estimators.

For  the noiseless  Mackey-Glass system,  it  can be  seen in  Fig.~\ref{fig:5.1} that  the MR  of the  first classifier
increases with $m$,  achieving its maximum ($0.065$) for $ m=19$.  Further, the second classifier presents a non-zero MR
only for  $m\ge15$.  The MR for  the third classifier  is zero for  $0\leq m\leq14$ with  a maximum value  of $0.006$ at
$m=15$.  A similar  behavior can be  observed for the  Mackey-Glass model  immersed in  noise (\textit{SNR}$=5$~dB).  In
contrast with  the noiseless case,  in Fig.~\ref{fig:5.2}  it is shown that  the MR  of the  classifier which  uses only
$ApEn_{max}$ has been greatly increased.  Additionally,  using only $r_{max}$, the classifier has non-zero MR for $2\leq
m\leq4$.  Nevertheless, the MR of the classifier that uses both estimators still remains equal to zero for $2\leq m \leq
4$ values.  For the  case in which the \textit{SNR}$=0$~dB  (Fig.~\ref{fig:5.3}),  it can be noticed that  the MR of the
third classifier is always below  or equal to the lowest MR between the other  two classifiers.  The last results attest
that, as an ensemble, $ApEn_{max}$ and $r_{max}$ provide features that are robust against noise.

Regarding the results for the noiseless Shilnikov's model, it is shown in Fig.~\ref{fig:5.4} that, for $3\leq m \leq 11$
the MR of the  first classifier is lower than the  second classifier's MR.  For $m=2$ as well as  for $12\leq m \leq20$,
the last statement is reversed.  However,  the  MR for the third classifier remains below the MR  of the other two ones,
being zero for $4\leq m  \leq 11$ and for $16\leq m \leq 20$.  Additionally,  from  Fig.~\ref{fig:5.5} it can be noticed
that for all $m$ values the MR of the third classifier is always below or equal to the lowest MR value between the other
two classifiers,  being zero for  $8\leq m\leq10$ and $0.004$ for $m=12$.  For the  \textit{SNR}$=0$~dB case,  it can be
observed in  Fig.~\ref{fig:5.6} that for  $m\ge8$ very low MR  values are achieved  by the first  and third classifiers,
being zero for $10\leq m \leq 15$.

Very similar results were achieved with the logistic map (Figs.~\ref{fig:5.7}, \ref{fig:5.8}, \ref{fig:5.9}).  The MR of
the classifier that uses in conjunction $ApEn_{max}$ and $r_{max}$ is  zero for all $m$ values in the noiseless case and
for $m\ge3$ in the \textit{SNR}=$5$~dB case.  From Fig.~\ref{fig:5.9} it can be seen that the MR of the third classifier
is below the  MR of the other classifiers  for all $m$ values,  being zero  for $m=8$ and $m=9$ and  achieves a very low
value for $m\ge5$.  This results lead us to think about the usefulness of these estimators to discriminate dynamics from
discrete-time non-linear systems.

It is important to notice that when these three  systems were immersed in high levels of noise (see Figs.~\ref{fig:5.3},
\ref{fig:5.6} and \ref{fig:5.9}) the  worst  results  were  achieved  for  low  $m$  values (specially for $m=2$).  This
suggests that increasing the  embedding dimension could be beneficial for the  discrimination process.  As a conclusion,
these results highlight the complementary relationship between both  estimators and the benefits of being used together.
It is also important to observe that the use of both estimators enlarges the range of $m$ values that can be selected to
achieve a good classification performance in presence of noise.  Nonetheless, using an estimate of the minimum embedding
dimension can be a wise choice (see Fig.~\ref{fig:5.3} and \ref{fig:5.6} for $m=12$).

There is an interesting fact in these results concerning the  presence of noise in the time series.  As it was discussed
before,  the addition of  noise not only decreases the  distance of $ApEn_{max}$ and $r_{max}$  curves between different
dynamics but  it also reduces  both estimators' CI.  The trade-off  between these two  phenomena is more  evident as the
\textit{SNR} is reduced.  In Figs.~\ref{fig:5.2} and ~\ref{fig:5.3}, it can be seen that for $m\ge4$ the MR of the first
classifier is  larger for \textit{SNR}$=5$~dB  than for \textit{SNR}$=0$~dB.  As  a consequence of  this trade-off,  the
distributions of $ApEn_{max}$ values for two different dynamics are more overlapped in the case with \textit{SNR}$=5$~dB
than  when  \textit{SNR}$=0$~dB.   The   last  statement  can  be  verified   comparing  the  Bhattacharyya  coefficient
(\textit{$B_{c}$})  \cite{Kailath1967}  between  the $ApEn_{max}$  distributions  of  different  dynamics  for different
\textit{SNR}s.  For two density functions  $p$  and  $q$  over  the  same  domain  $X$,  this  coefficient is defined as
$B_{c}\left(p,q\right)=\sum_{x\in X}\sqrt{p(x)q(x)}$,  $0  \leq B_{c}\leq 1$,  being zero  if $p(x)$  and $q(x)$  do not
overlap.   The  $B_{c}$   coefficients  between   $ApEn_{max}$  distributions   ($m=19$)  for   \textit{SNR}$=5$~dB  and
\textit{SNR}=$0$~dB are $0.91$ and  $0.53$ respectively.  This fact explains why the MR of  this classifier is lower for
\textit{SNR}$=0$~dB  than for  \textit{SNR}$=5$~dB when  high $m$  values are  used with  systems like  Mackey-Glass and
Shilnikov's.

\begin{figure}[!t]
    \hspace{-1.6cm}{
    \subfloat[][]{\includegraphics[width=0.390\linewidth,height=\textheight,keepaspectratio]{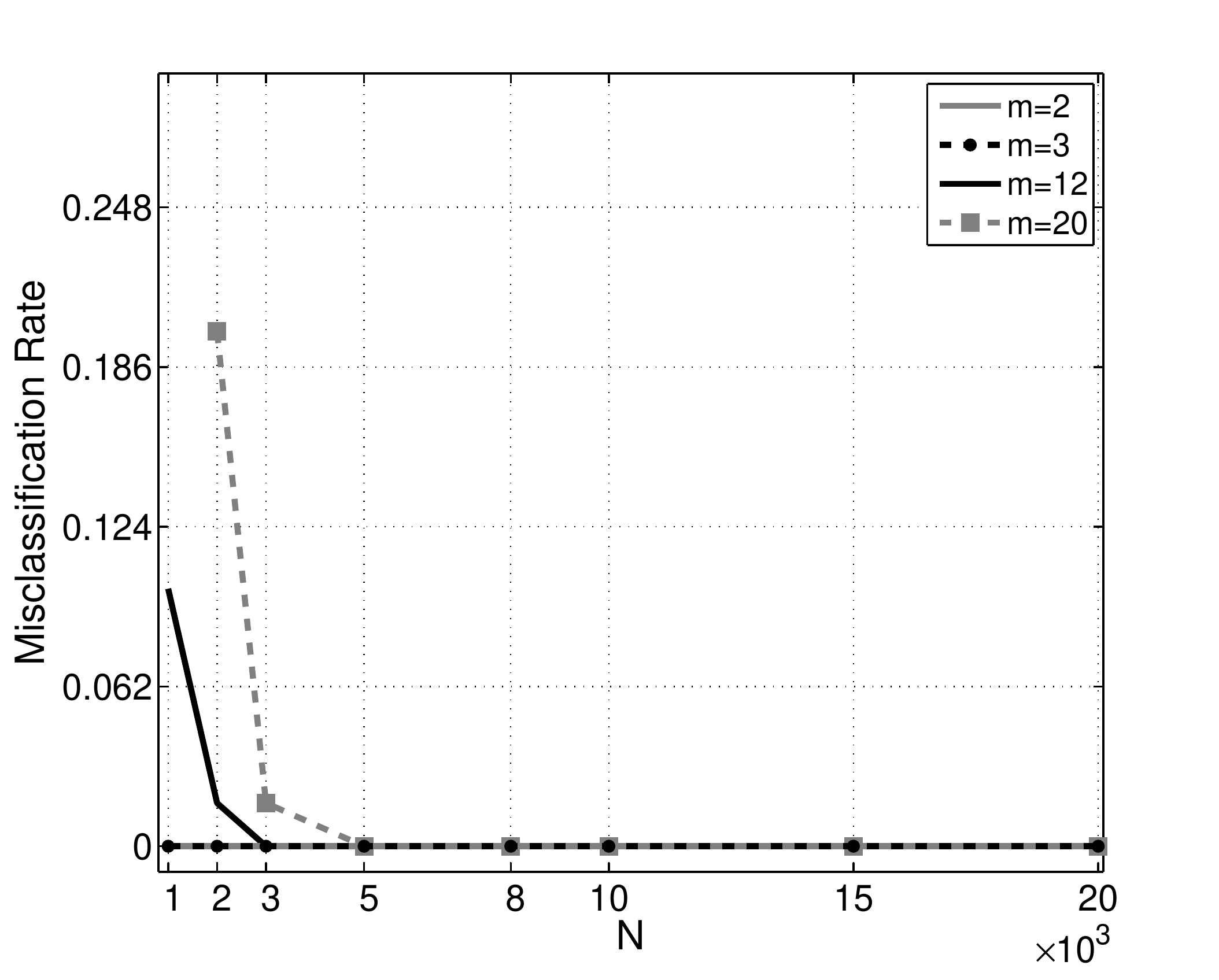}\label{fig:6.1}}\hspace{-5mm}
    \subfloat[][]{\includegraphics[width=0.390\linewidth,height=\textheight,keepaspectratio]{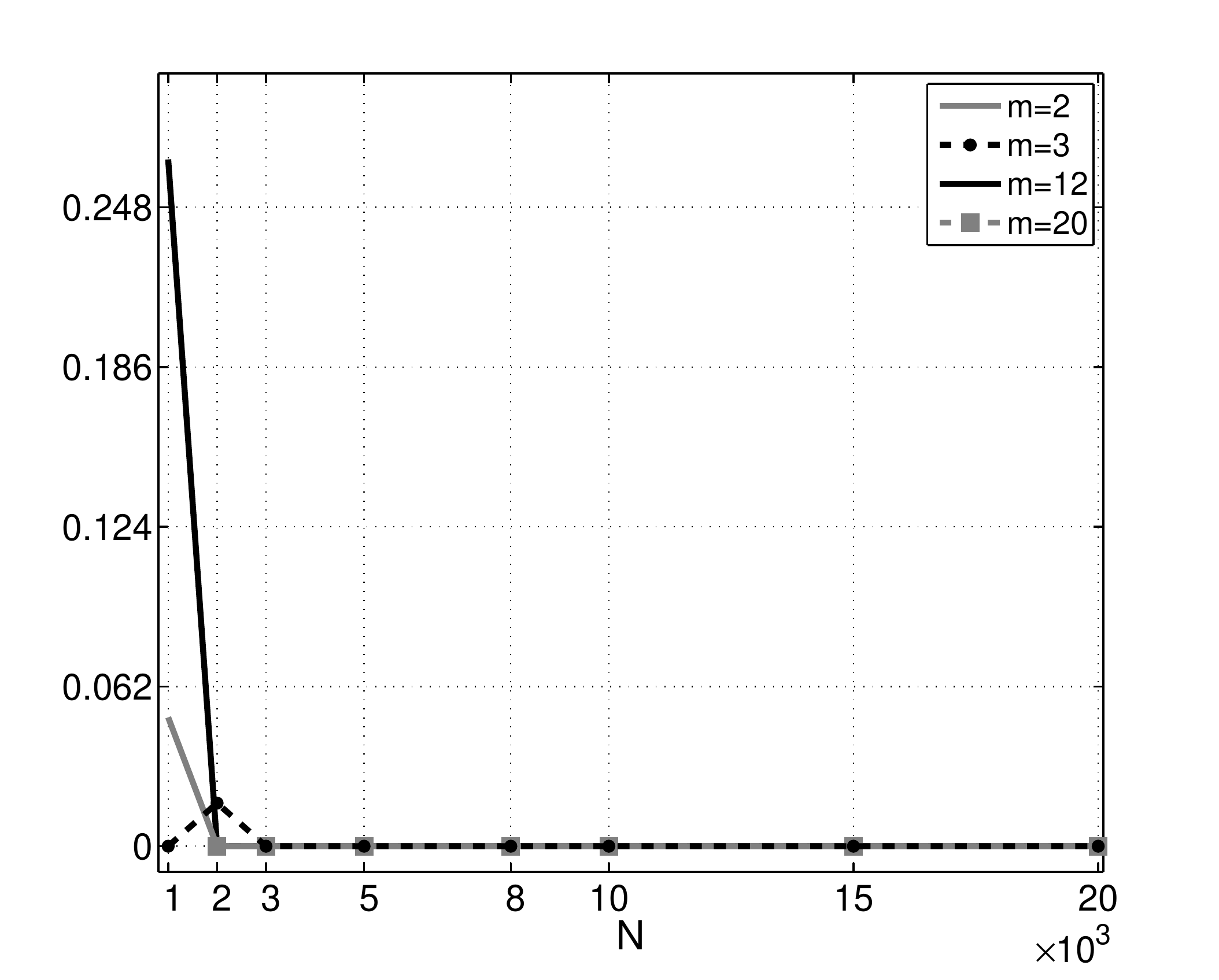}\label{fig:6.2}}\hspace{-5mm}
    \subfloat[][]{\includegraphics[width=0.390\linewidth,height=\textheight,keepaspectratio]{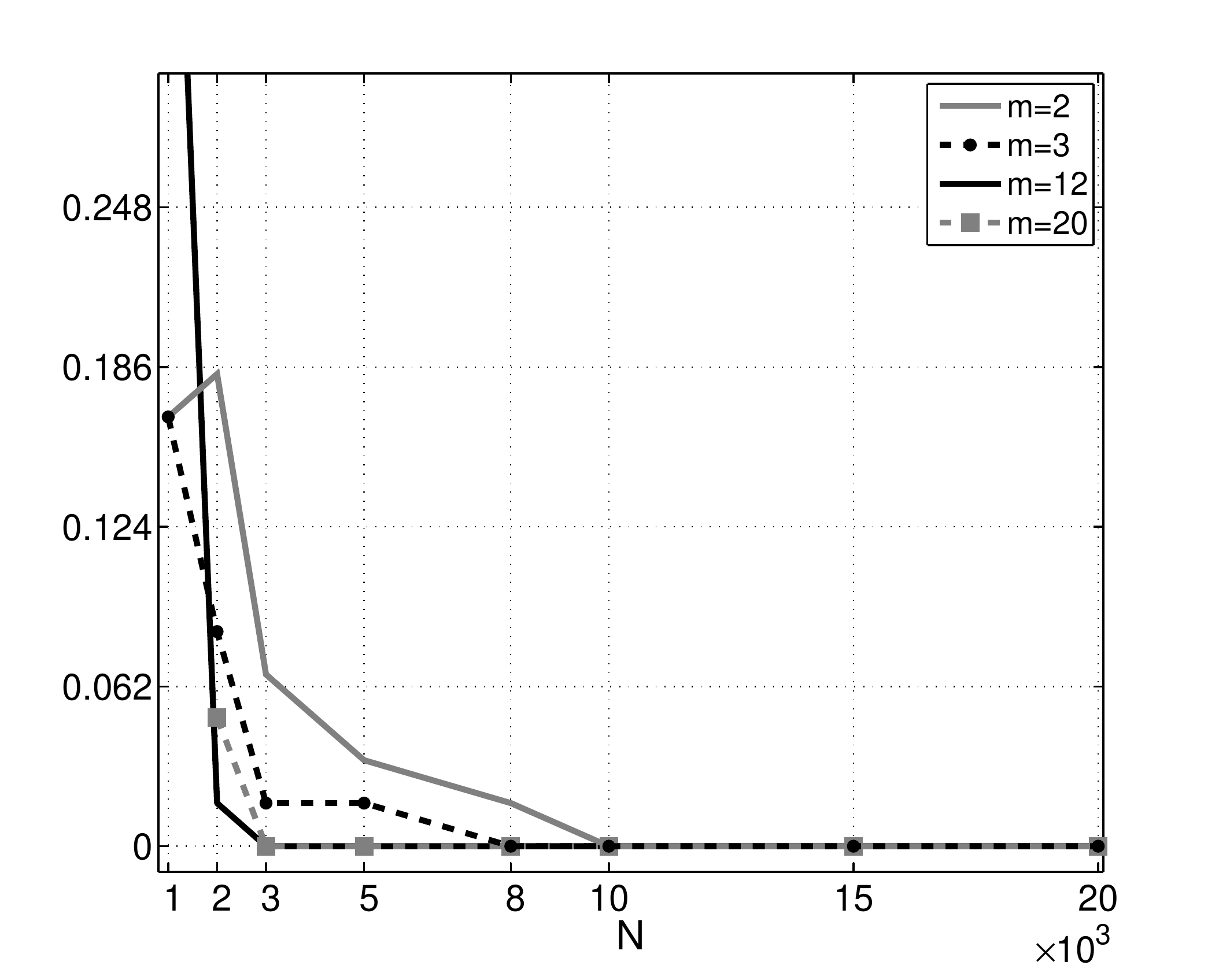}\label{fig:6.3}}\\}
    \vspace{-5mm}
    \hspace{-1.6cm}{
    \subfloat[][]{\includegraphics[width=0.390\linewidth,height=\textheight,keepaspectratio]{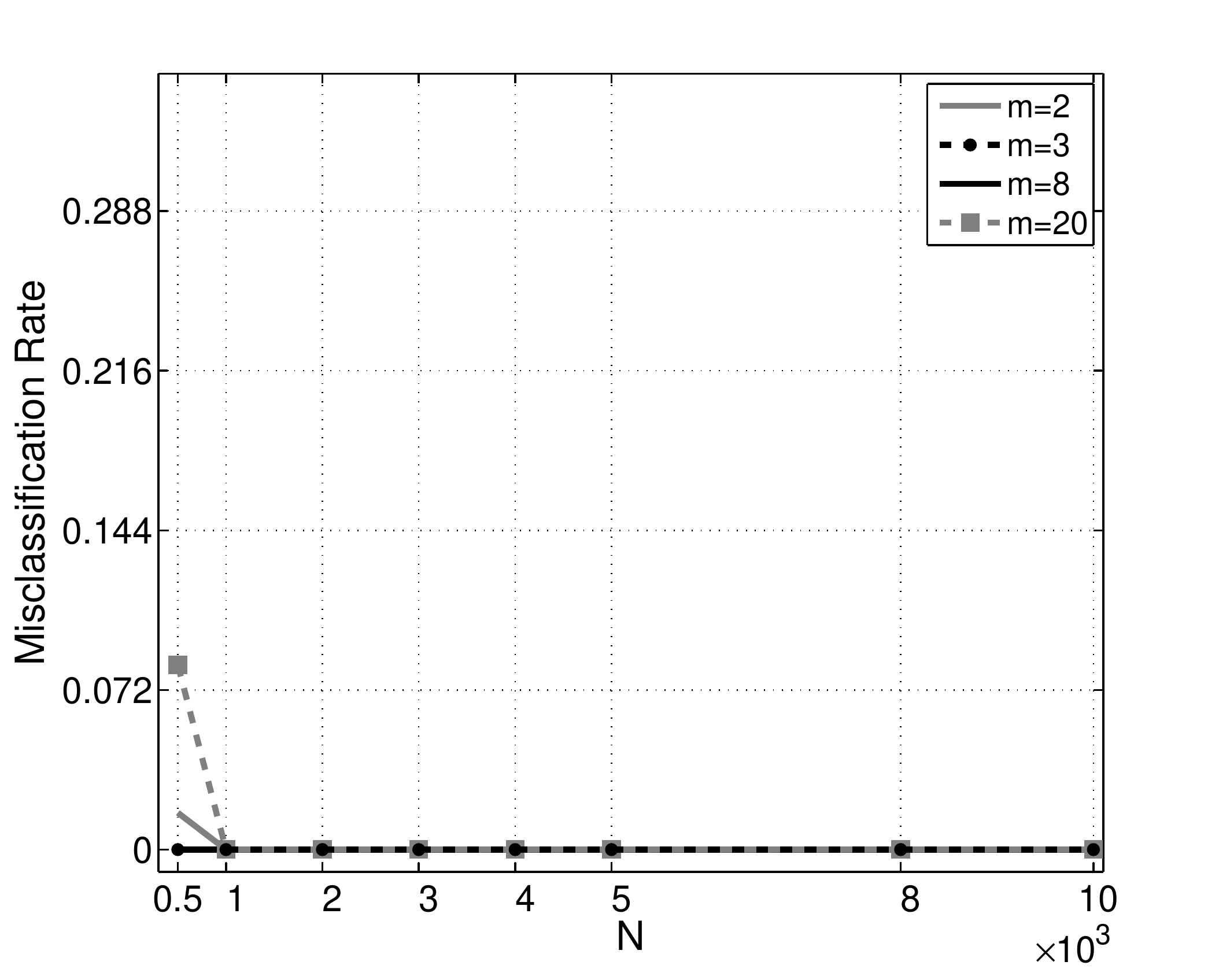}\label{fig:6.4}}\hspace{-5mm}
    \subfloat[][]{\includegraphics[width=0.390\linewidth,height=\textheight,keepaspectratio]{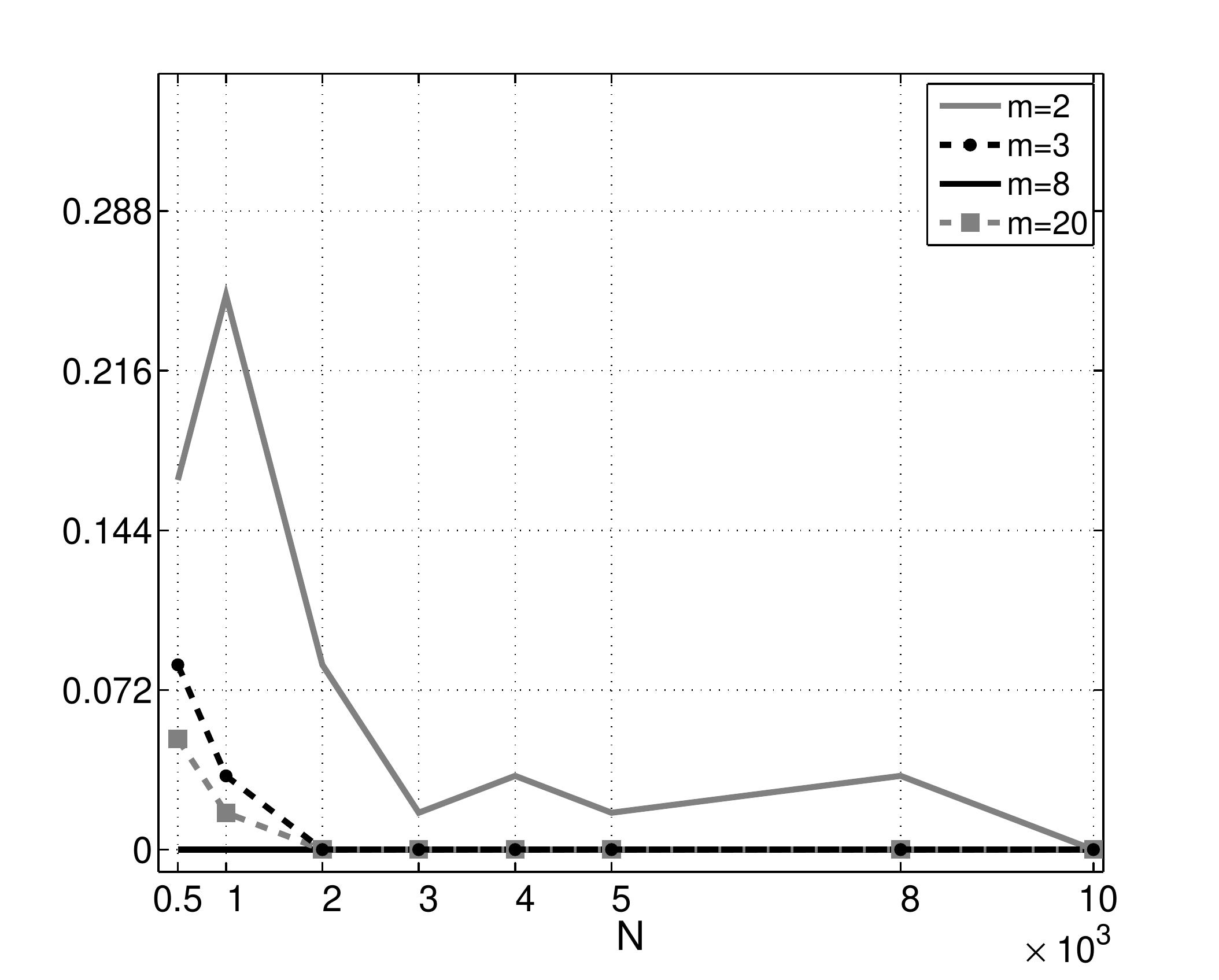}\label{fig:6.5}}\hspace{-5mm}
    \subfloat[][]{\includegraphics[width=0.390\linewidth,height=\textheight,keepaspectratio]{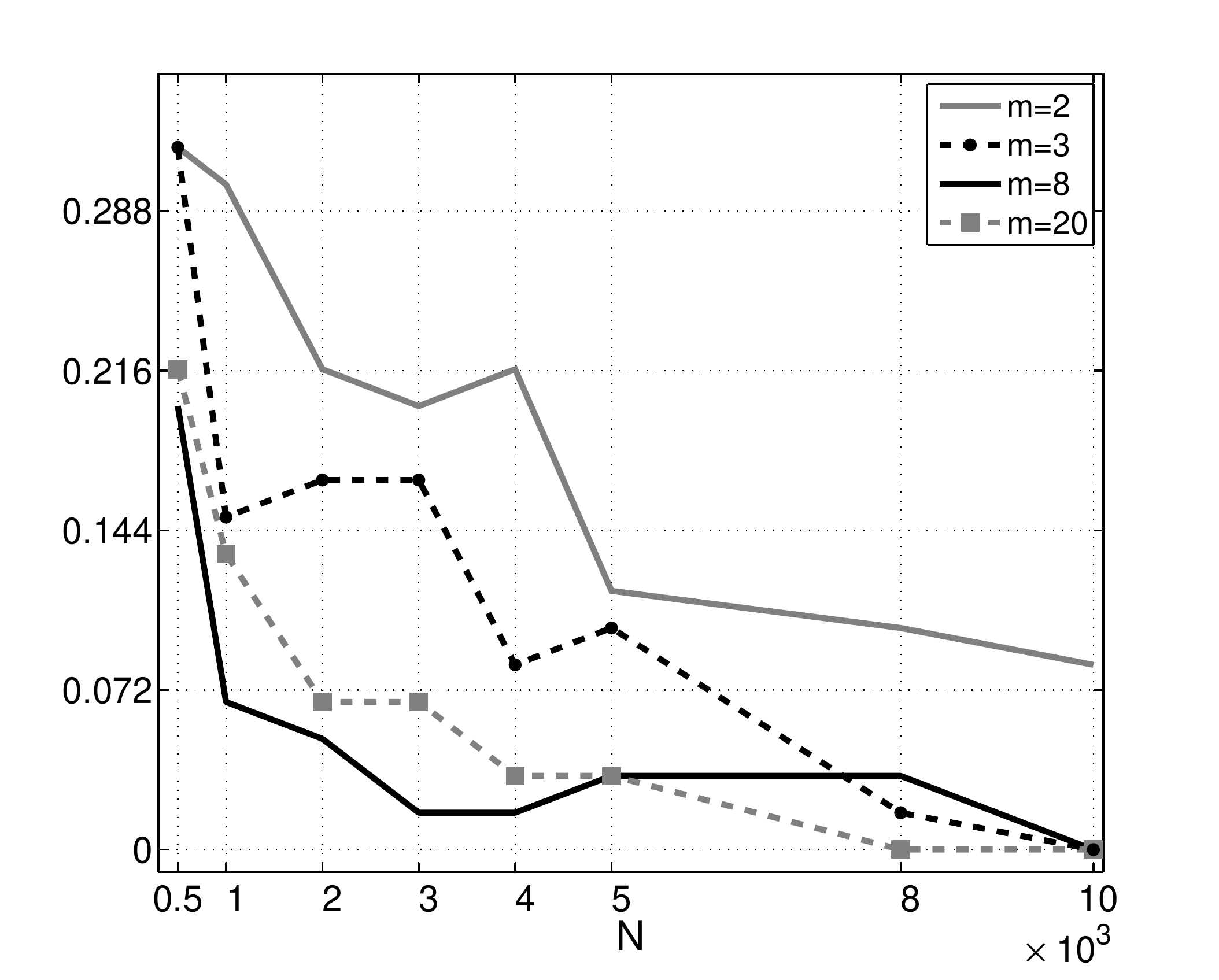}\label{fig:6.6}}\\}
    \caption{Misclassification  rate   as  a  function   of  the  data   length.   Mackey-Glass  model:   (a)~noiseless.
    (b)~\textit{SNR}$=5$~dB.    (c)~\textit{SNR}$=0$~dB.    Logistic   map:   (d)~noiseless.    (e)~\textit{SNR}$=5$~dB.
    (f)~\textit{SNR}$=0$~dB.}
    \label{fig:6}
\end{figure}

An important topic  that must be considered in  the calculation of $ApEn_{max}$ and $r_{max}$  is the data length.  When
the time series is short, the choice of large $m$ and $\tau$ values can be harmful because the estimation of conditional
probabilities becomes  unreliable \cite{Pincus1991,Pincus1994}.  However,  there is  another issue that  can alter their
estimation and it is related to the use of small $m$ values.

It is known  that poor state space reconstructions  are obtained when the system  is embedded with an  $m$ value smaller
than  the system's  minimum embedding  dimension,  and such  situation  brings  to  the  occurrence  of  false neighbors
\cite{Wolf1985},  i.e.  points that  are close  due to a  low embedding dimension  rather than  because of  the system's
dynamics.  Given  that the  estimation of  conditional probabilities  is based  on counting  neighbors' occurrences,  an
appropriate selection of  the $m$ value demands  to take into account  an estimation of the  minimum embedding dimension
\cite{Small2005}.  With the  aim of  assess the  behavior of  our method  as a  function of  the data  length,  the next
simulation was conducted over the Logistic map and the Mackey-Glass system.

Two sets of $30$  realizations were built.  Each set was obtained  using a different value of the  $R$ parameter for the
Logistic map and,  of the $c$  parameter for the Mackey-Glass system.  Each signal of these  sets was normalized to have
unitary energy.  For the Logistic map $ApEn_{max}$ and $r_{max}$ were estimated with $m=\left[2,3,8,20\right]$, $\tau=1$
and $N=\left[0.5,1,2,3,4,5,8,10\right]\times10^{3}$.  For the  Mackey-Glass system these estimators  were evaluated with
$m=\left[2,3,12,20\right]$,  $\tau=83$ and $N=\left[1,2,3,5,8,10,15,20\right]\times10^{3}$.  Then, the misclassification
rate of a linear SVM  classifier  that  uses  both  estimators  as  features  was  computed using \textit{Leave one Out}
cross-validation.  Additionally,  the same  procedure was used  over the same  signals contaminated with  white Gaussian
additive noise (\textit{SNR}$=5$~dB and  \textit{SNR}$=0$~dB).  It is important to mention that the  values of $m=8$ and
$m=12$ were  suggested by  the Cao's  algorithm \cite{Cao1997}  as minimum  embedding dimensions  for the  noisy signals
(\textit{SNR}$=0$~dB) from the Logistic map and, from the Mackey-Glass system respectively.  For the Mackey-Glass system
the calculations with $m=20$ were made for all $N$ values except $N=1000$.

In Fig.~\ref{fig:6}  it is shown  the misclassification rate calculated  for both systems as  a function of  $N$ and the
noise level.  In Fig.~\ref{fig:6.1}  are shown the results for  the noise free Mackey-Glass system.  It  can be observed
that for small data length values the biggest errors are achieved using $m=20$ ($N=2000$) and $m=12$ ($N=1000$); this is
a consequence of the  reduced amount of information available to  estimate the conditional probabilities.  Nevertheless,
as $N$ is increased,  the error for all $m$ values goes to  zero.  It must be noticed that for $m=2$ and $m=3$ the error
is zero for all $N$ values.

On the other hand Fig.~\ref{fig:6.2} shows that, compared with the noiseless case, for $m=2$ (at $N=1000$) and $m=3$ (at
$N=2000$) the error has increased its value from zero,  whereas the error for $m=12$ and $m=20$ has decreased its values
to  zero  for  $N=2000$.   Observe  that  the  error  is  zero  from  $N\ge3000$  regardless  the  value  of  $m$.  From
Fig.~\ref{fig:6.3} it can be seen that,  excluding the error (equal to $0.48$) for $m=12$ at $N=1000$, the biggest error
is accomplished using $m=2$ followed by the one obtained with $m=3$ for $1000\leq N \leq 8000$.  However,  the error for
$m=12$ and $m=20$ is always  lesser or equal to the error achieved with $m=2$  or $m=3$,  moreover,  it is zero starting
from $N=3000$.  Comparing Figs.~\ref{fig:6.1}  and \ref{fig:6.3} for m  = 2 and m =  3 it is clear  that,  for small $N$
values,  a poor state  space reconstruction added to the  presence of noise deteriorates the  discrimination capacity of
$ApEn_{max}$ and $r_{max}$.

For the  noiseless Logistic map  (Fig.~\ref{fig:6.4}) it can  be observed that  for $N=500$  the biggest  error ($0.08$)
belongs to the estimators  calculated with $m=20$ followed by the error  calculated with $m=2$ ($0.017$).  However,  for
$m=3$ and $m=8$ the error is zero.  Moreover,  as $N$ is increased,  the error remains equal to zero for all $m$ values.
It can be  seen Fig.~\ref{fig:6.5} that the  biggest error is achieved with  $m=2$ for all $N$  values except $N=10000$.
Instead,  for $m=8$ the error is  equal to zero for all $N$ values.  It is worth mentioning  that for all $N$ values the
error obtained with $m=8$  and  $m=20$  is  always  below  or  equal  to  the  error  attain with $m=2$ and $m=3$.  From
Fig.~\ref{fig:6.6} it can  be noticed that using $m=2$  produces the worst classification error  regardless the value of
$N$ and the best results are accomplished using $m=8$ and $m=20$ for almost all $N$ values.

Based on  these results  we can conclude  that in the  $ApEn_{max}$ and  $r_{max}$ estimation's  processes it  is highly
recommended to keep in mind  that there exists a trade-off between $m$ and $N$,  and  special attention is needed in the
presence of noise.  When data length is short and there  is not noise in the signal,  relative small $m$ values provides
the best  performance.  However,  in presence of noise,  it  would be wise either  to use an estimation  of the system's
minimum embedding dimension whenever it is possible,  or to use a  value as close as possible to it when the data length
is a  limitation.  It must also  be considered that in  real applications,  such as  epileptic seizures' detection,  the
duration of  some events  is only  of a  few samples:  for example  absence seizures  often last  less than  $5$ seconds
\cite{Shorvon2004},  which correspond to  $1280$~samples using a standard sampling  frequency of $256$~Hz.  Although for
small $N$ values  there is no guarantee of an  accurate estimation of $ApEn_{max}$ nor $r_{max}$  with relative high $m$
values.  The results here presented show that using $m$ values above $2$ or $3$ can increase the discrimination capacity
of these estimators, specially in presence of noise.

\begin{figure}[!t]
    \centering
    \subfloat[][]{\includegraphics[width=0.500\linewidth,height=\textheight,keepaspectratio]{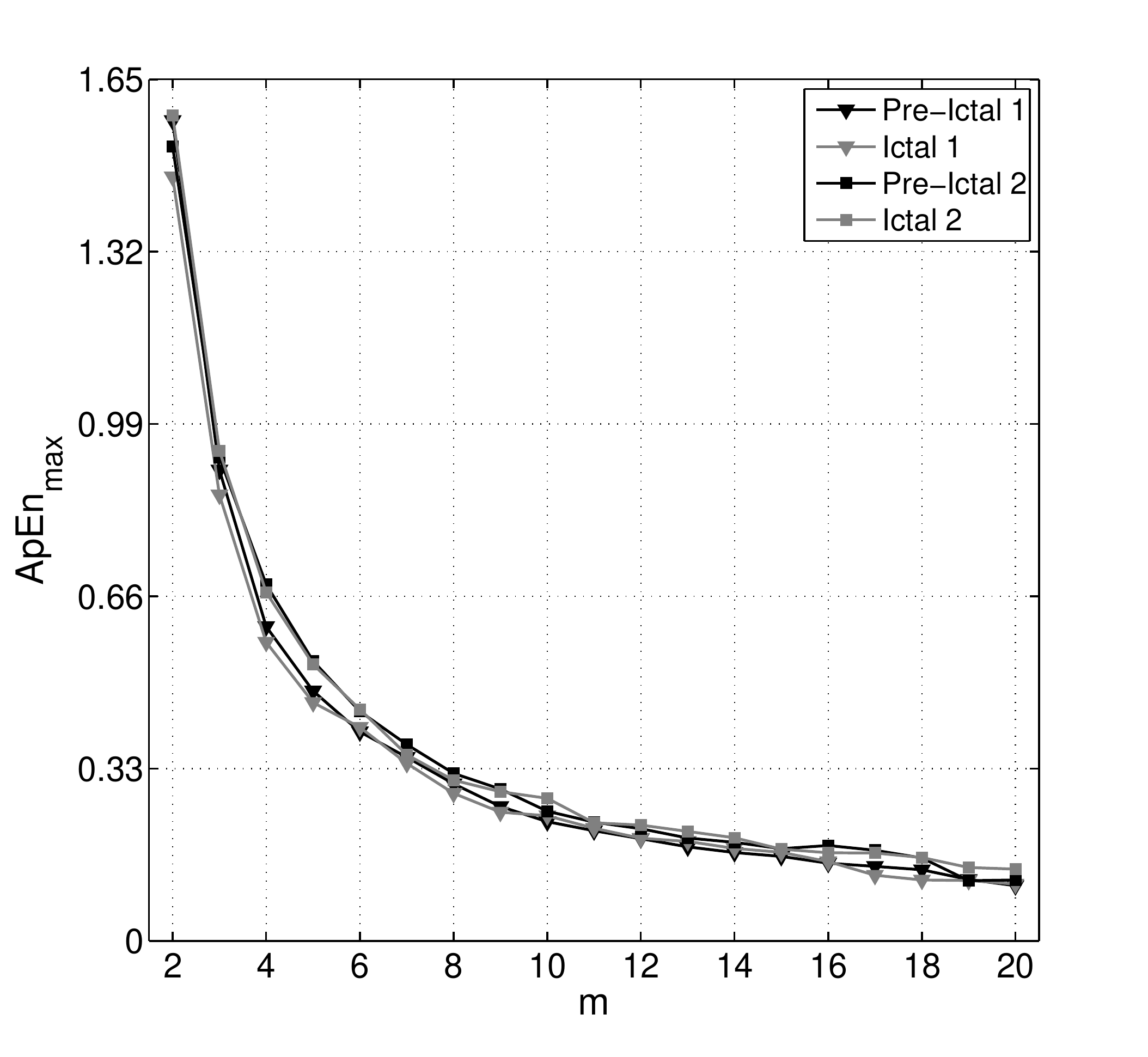}\label{fig:7.1}}\hspace{-5mm}
    \subfloat[][]{\includegraphics[width=0.500\linewidth,height=\textheight,keepaspectratio]{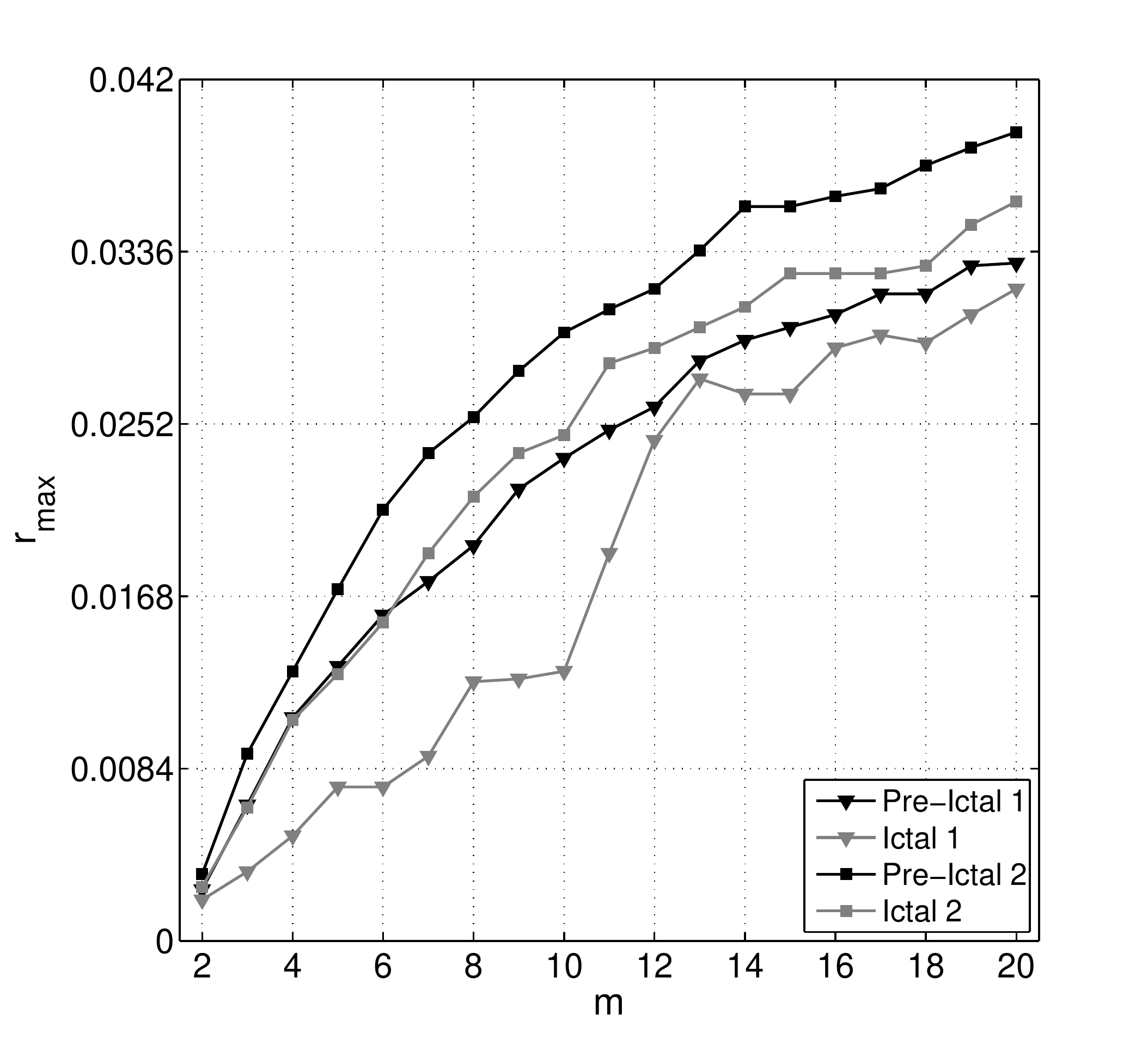}\label{fig:7.2}}
    \vspace{-2mm}
    \subfloat[][]{\includegraphics[width=0.500\linewidth,height=\textheight,keepaspectratio]{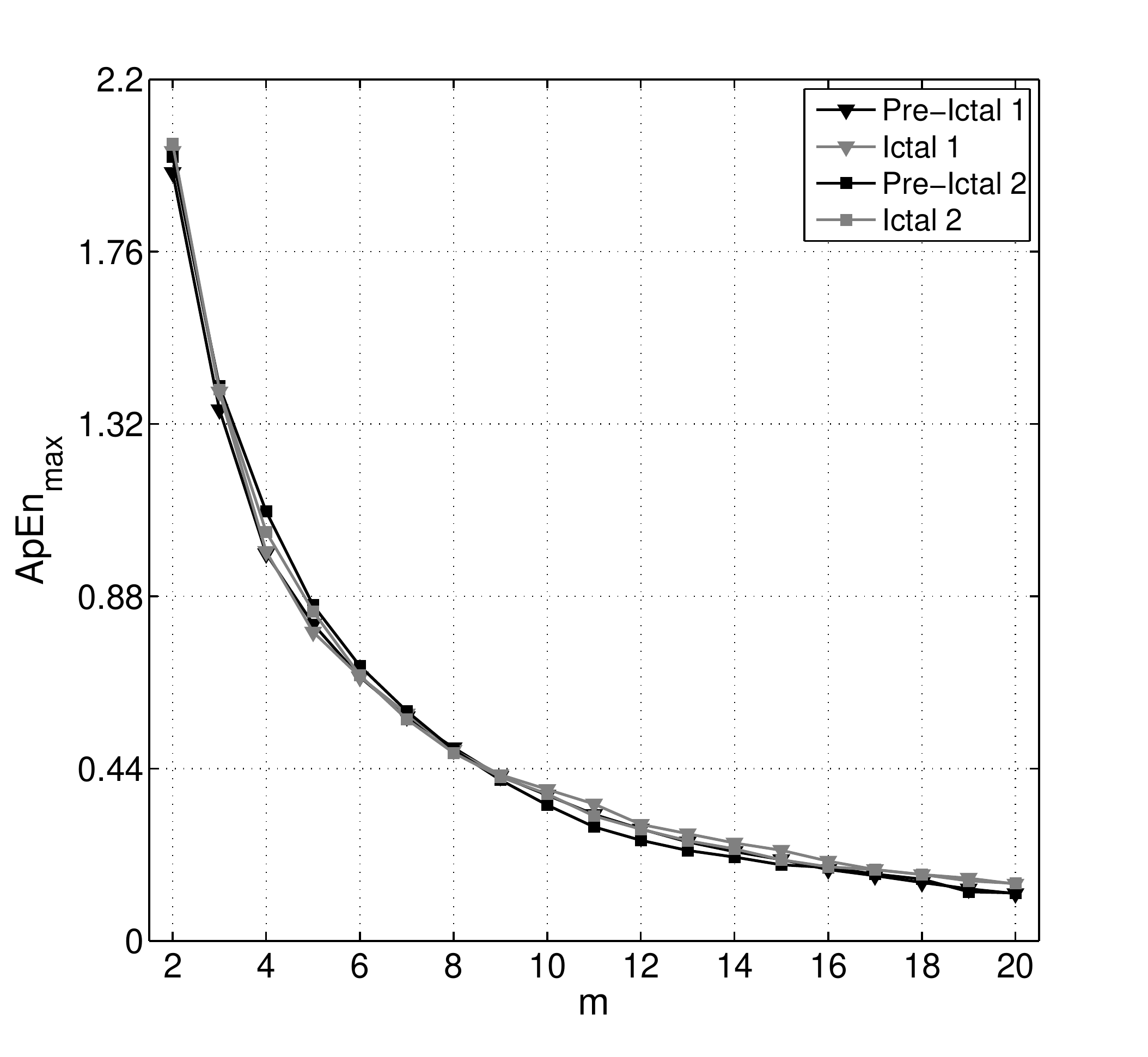}\label{fig:7.3}}\hspace{-5mm}
    \subfloat[][]{\includegraphics[width=0.500\linewidth,height=\textheight,keepaspectratio]{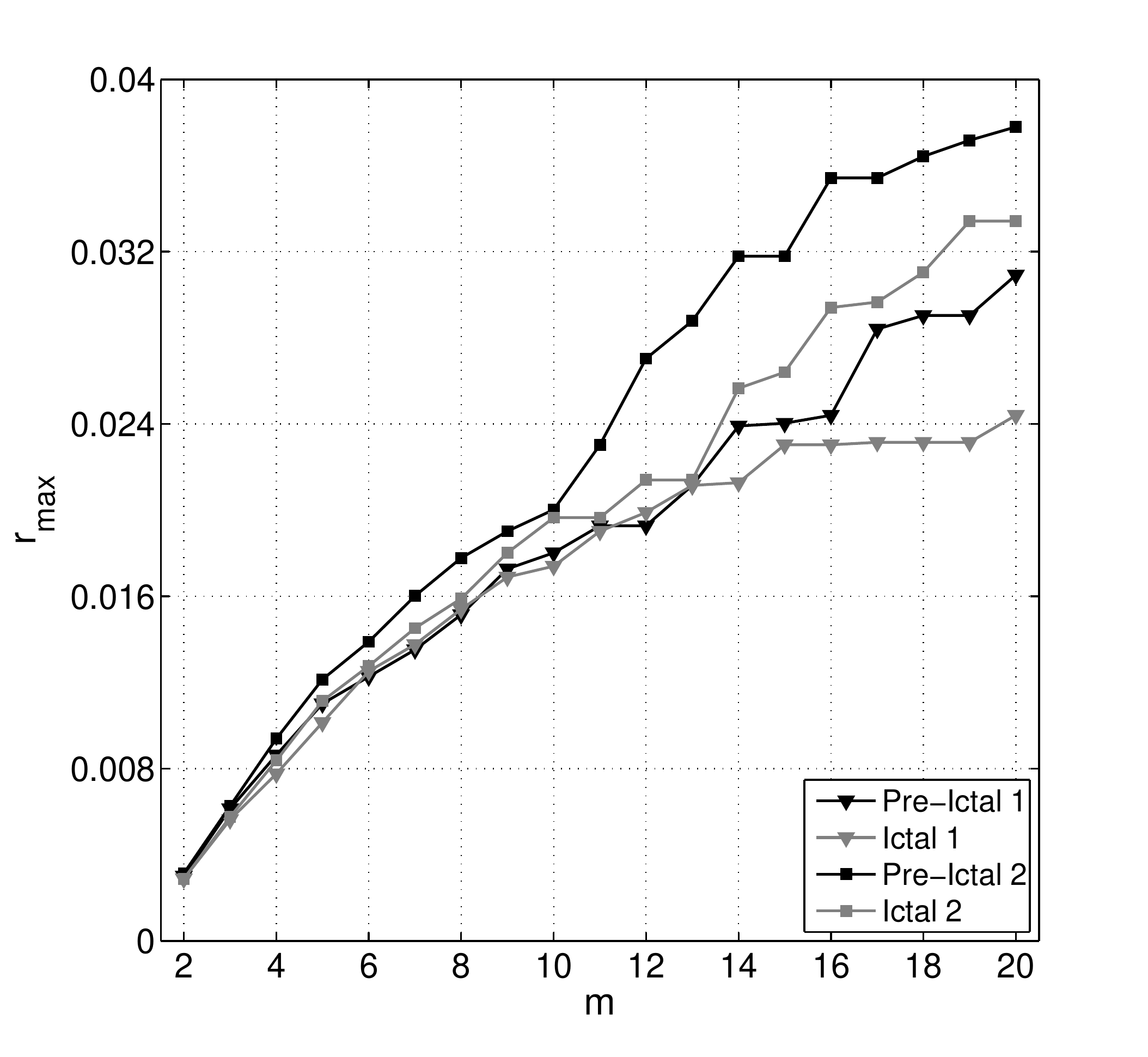}\label{fig:7.4}}
    \caption{EEG signal. Two ictal episodes and their respective pre-ictal segments. Raw signal: (a) $ApEn_{max}$. (b) $r_{max}$.
     \textit{SNR}$=5$~dB: (c) $ApEn_{max}$. (d) $r_{max}$.}
    \label{fig:7}
\end{figure}

The studies conducted on the EEG recording provided similar results to those obtained with the previous simulations.  In
Fig.~\ref{fig:7} are  presented the $ApEn_{max}$  and $r_{max}$ curves  as functions of  $m$ for two  ictal episodes and
their respective  pre-ictal segments.  The distances (relative  to the scale)  between the  $ApEn_{max}$ curves  of each
ictal  and  its  corresponding  pre-ictal  episodes  are  small  for  all  the  $m$  values,   as  can  be  observed  in
Fig.~\ref{fig:7.1}.  On the  contrary,  Fig.~\ref{fig:7.2} suggests that $r_{max}$  can be used  to discriminate between
dynamics.  Decreasing the  \textit{SNR} tends  to reduce  the distance  between $ApEn_{max}$  and $r_{max}$  curves (see
Figs.~\ref{fig:7.3} and \ref{fig:7.4}).  However,  for high $m$ values, the information given by $r_{max}$ can be useful
to distinguish between dynamics.

\begin{figure}[!t]
     \centering
     \subfloat[][]{\includegraphics[width=1.000\linewidth,height=\textheight,keepaspectratio]{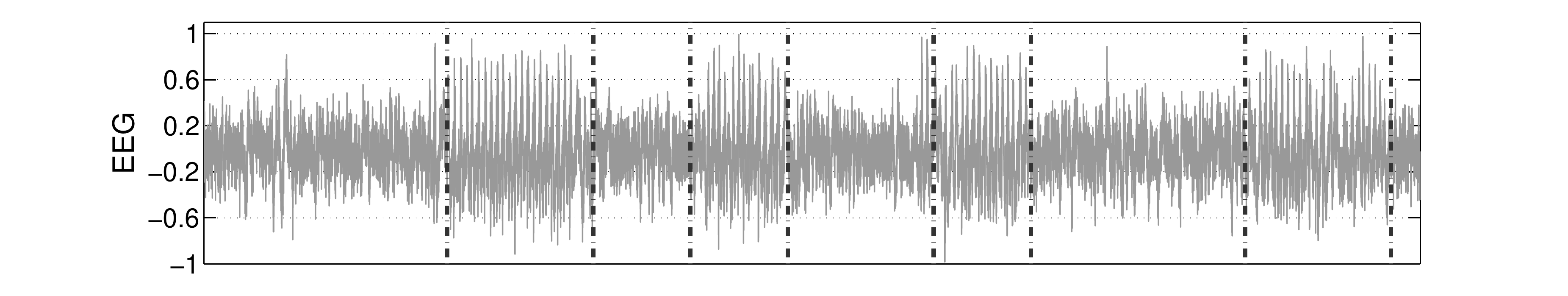}\label{fig:8.1}}
     \vspace{-6.5mm}      
     \subfloat[][]{\includegraphics[width=1.000\linewidth,height=\textheight,keepaspectratio]{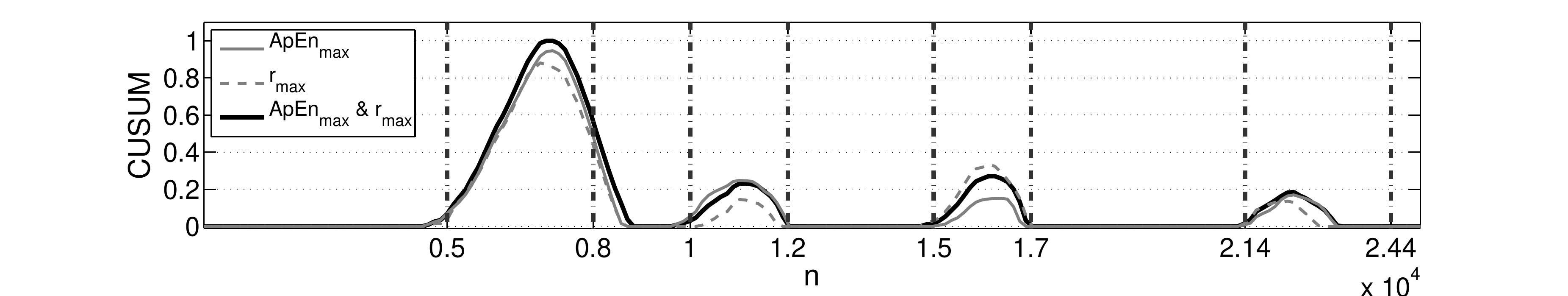}\label{fig:8.2}}
     \caption{Ictal  episodes  detection using  $ApEn_{max}$  and $r_{max}$  from  an  EEG  signal  with  additive noise
     (\textit{SNR}$=5$~dB):  (a) EEG signal.  (b)  CUSUM  over  the  first  PC  on the $\mat{C^{A}}$,  $\mat{C^{R}}$ and
     $\mat{C^{AR}}$ matrices.  The ictal episodes can be found between vertical dashed lines.}
     \label{fig:8}
\end{figure}

It is  worth mentioning that,  for  this signal and with  these estimators,  it is  difficult to separate  the ictal and
pre-ictal episodes as isolated  groups.  However,  it is possible to state differences between  an ictal episode and its
corresponding pre-ictal one.  This result leads  us to think that a suitable approach to  detect ictal episodes from EEG
signals, using these estimators, should be one in which their temporal evolution could be evaluated.

In order  to assess  this idea,  we  corrupted the EEG  signal with  white Gaussian  noise (\textit{SNR}$=5$~dB)  and we
considered sliding windows of  length $N=1000$,  shifted $128$ data points.  Each window was  normalized to have unitary
energy.  $ApEn_{max}$ and $r_{max}$ were estimated using $\tau=10$ for  $2\leq m \leq 6$.  With these results we proceed
as follows: first, we built the matrices $\mat{C^{A}}$ and $\mat{C^{R}}$, where the entry $C^{A}_{k,i}$ was the value of
$ApEn_{max}$ calculated with the \textit{i}-th value of  $m$ for the \textit{k}-th window.  The matrix $\mat{C^{R}}$ was
built alike with the $r_{max}$ values.  Each matrix was  statistically normalized (zero mean and unitary \textit{SD}) by
columns.  Observe that the temporal evolution of $ApEn_{max}$  and $r_{max}$ calculated with the \textit{i}-th $m$ value
can be  evaluated by  looking the  \textit{i}-th column of  the $\mat{C^{A}}$  and $\mat{C^{R}}$  matrices respectively.
A third  matrix named  $\mat{C^{AR}}$ was conformed  by the horizontal  concatenation of  the above  mentioned matrices:
$\mat{C^{AR}}=\left(\mat{C^{A}}\,|\,\mat{C^{R}}\right)$.   Next,   we   performed   a   Principal   Component   Analysis
(\textit{PCA}) over  each matrix.  The  first principal component  ($1^{st}$PC) can be  thought as  a summary  that best
represents the information collected by these estimators through all $m$ values.  Finally, an algorithm for detection of
abrupt mean  changes (\textit{CUSUM}) was  applied on the  $1^{st}$PC of each  matrix \cite{Montgomery2010}.  The target
mean and the reference value were fixed as the average  of the first twenty data points and two times their \textit{SD},
respectively.

In Fig.~\ref{fig:8}  are presented the  EEG signal with  the four ictal  episodes marked  between vertical  dashed lines
(Fig.~\ref{fig:8.1}) and the  results  of  the  \textit{CUSUM}  algorithm  applied  over  the  $1^{st}$PC of each matrix
(Fig.~\ref{fig:8.2}).  It  can be  observed in  Fig.~\ref{fig:8.2} that  all ictal  episodes can  be detected  using the
information contained in  each matrix.  Nevertheless,  while some ictal episodes are  better detected with $ApEn_{max}$,
others are better detected with $r_{max}$.  On the other hand, a more consistent identification of ictal episodes can be
achieved using  in conjunction  both estimators.  These  results suggest  that,  with the  information provided  by both
$ApEn_{max}$ and  $r_{max}$ for different $m$  values,  the ability to  discriminate between  different dynamics  can be
increased  (even in  presence of  noise),  since changes  that cannot  be identified  in the  temporal evolution  of one
estimator could be identified in the temporal evolution of  the other one.  It must be remarked that these findings only
suggest the suitability to  jointly use both estimators to detect ictal  episodes from EEG signals.  Further experiments
with a large data base will be conducted in future  works to statistically assess the performance of the proposed method
to detect complexity changes in real signals.

\section{Conclusions}
\label{sec:con}
The approximate  entropy has  been recognized by  its ability to  distinguish between  different system’s  dynamics when
short-length data with  moderate noise are available.  However,  it is  also known that high noise  levels and incorrect
parameter selection can undermine its discrimination capacity.  In  order to overcome these difficulties,  in this paper
we have proposed a method  based on the use of $r_{max}$ along with $ApEn_{max}$  to discern between different dynamics.
Using signals from real physiological and from simulated low- and high-dimensional systems,  with and without noise,  we
have studied the behavior of $ApEn_{max}$ and $r_{max}$ as functions of the embedding dimension, the data length and the
noise level.  The results indicate that,  even in presence of noise, $r_{max}$ provides valuable information that can be
used for classification purposes.  Furthermore, as these estimators vary with $m$, there is a complementary relationship
between them,  which strengthens the idea of using $ApEn_{max}$ combined with $r_{max}$ to distinguish between dynamics.
Cross-validation simulations have demonstrated that the jointly use of both estimators as input features,  significantly
decreases the  misclassification rate of  a simple linear  classifier.  Moreover,  the conjoint  use of  both estimators
enlarges the range of $m$  values that can be chosen to achieve a  good classification performance.  Concerning the data
length,  we have shown that  for short-length signals good discriminating features can  be achieved using relative small
$m$ values if there  is  no  noise.  However,  in  presence  of  noise  the  discrimination capacity of $ApEn_{max}$ and
$r_{max}$ can be  increased using $m$ values above  $2$ or $3$.  Our results encourage  the use of an  estimation of the
system's minimum embedding dimension  when it is possible,  or the use  of a close enough value when  the data length is
a limitation.  We assert that as well as $ApEn_{max}$,  the  estimator $r_{max}$ can also be utilized to discern between
dynamics even  in the presence  of noise.  Moreover,  the use of  $r_{max}$ has shown to  be helpful in  such cases when
$ApEn_{max}$ is unable to contrast between processes that  are immersed in noise.  The link between $r_{max}$ and system
complexity will be addressed in future studies, to reveal the nature of this relationship and its physical meaning.
\section*{Acknowledgments}
\label{sec:ack}
\small{
This work was supported by the National Agency for Scientific and Technological Promotion (ANPCyT), Universidad Nacional
de Entre Ríos, and the National Scientific and Technical Research Council (CONICET) of Argentina.
}
\bibliographystyle{elsarticle-num}
\bibliography{ApEnPhyA}
\end{document}